\documentclass[letterpaper, 10 pt, conference,dvipdfmx]{ieeeconf}

\IEEEoverridecommandlockouts                              % This command is only needed if 
\usepackage{graphics} % for pdf, bitmapped graphics files
\usepackage{epsfig} % for postscript graphics files
\usepackage{amsmath} % assumes amsmath package installed
\usepackage{amssymb}  % assumes amsmath package installed

\usepackage{color}
\usepackage{xcolor}
\usepackage{subfiles}
\usepackage{comment}
\usepackage{bm}
\usepackage{subcaption}

\newtheorem{theorem}{Theorem}

\newtheorem{lemma}{Lemma}
\newtheorem{remark}{Remark}

\newtheorem{definition}{Definition}

\newcommand{\mbr}{\mathbb{R}}
\newcommand{\red}[1]{\textcolor{red}{#1}}

\newcommand{\violet}[1]{\textcolor{violet}{#1}}

\title{\LARGE \bf
Design and Control of a VTOL Aerial Vehicle Tilting its Rotors Only with Rotor Thrusts and a Passive Joint
}

\author{Takumi Ito$^{1}$, Riku Funada$^{1}$, and Mitsuji Sampei$^{1}$% <-this % stops a space
\thanks{\copyright 2024 IEEE.  Personal use of this material is permitted.  Permission from IEEE must be obtained for all other uses, in any current or future media, including reprinting/republishing this material for advertising or promotional purposes, creating new collective works, for resale or redistribution to servers or lists, or reuse of any copyrighted component of this work in other works.}% <-this % stops a space
\thanks{*This work was supported by JSPS KAKENHI, Grant Number 21H01348 and JST SPRING Grant Number JPMJSP2106.}% <-this % stops a space
\thanks{$^{1}$ Department of Systems and Control Engineering, Tokyo Institute of Technology, 
        {\tt\small  takumi.ito@sl.sc.e.titech.ac.jp.} {\tt\small\{funada\},\{sampei\}@sc.e.titech.ac.jp.}}%
}

\begin{document}

\maketitle
\thispagestyle{empty}
\pagestyle{empty}

%%%%%%%%%%%%%%%%%%%%%%%%%%%%%%%%%%%%%%%%%%%%%%%%%%%%%%%%%%%%%%%%%%%%%%%%%%%%%%%%
% \documentclass[ieeeconf]{subfiles}
% \begin{document}
\begin{abstract}
This paper presents a novel VTOL UAV that owns a link connecting four rotors and a fuselage by a passive joint, allowing the control of the rotor's tilting angle by adjusting the rotors' thrust. This unique structure contributes to eliminating additional actuators, such as servo motors, to control the tilting angles of rotors, resulting in the UAV's weight lighter and simpler structure. We first derive the dynamical model of the newly designed UAV and analyze its controllability.
%Then, we present the control strategy for all flight modes, where \blue{ロータ傾斜角を変化させることで，aerodynamicsに関係する迎角と前方への速度を重点的に制御し，transition中の予期せぬダイナミクスの変化を抑えることができる}.
Then, we design the controller that leverages the tiltable link with four rotors to accelerate the UAV while suppressing a deviation of the UAV's angle of attack from the desired value to restrain the change of the aerodynamic force. 
Finally, the validity of the proposed control strategy is evaluated in simulation study.
%This paper desines a novel VTOL aerial vehicle, which 

%This paper introduces a novel concept for a tiltrotor VTOL aerial vehicle. The vehicle comprises rotors and a link connected to the fuselage via a passive joint. The uniqueness of this structure lies in its reliance solely on rotor thrust to control both the rigid-body vehicle and the tilt angle. This design reduces the weight of the tilting mechanism and enhances flexibility in tilt angle control. The contributions of this paper are threefold:
%1) modeling and controllability analysis of the new vehicle, 2) control strategy and controller proposal, 3) verifying the controllability of the system and assessing the utility of the controller through simulation experiments.

\end{abstract}

% \end{document}

%%%%%%%%%%%%%%%%%%%%%%%%%%%%%%%%%%%%%%%%%%%%%%%%%%%%%%%%%%%%%%%%%%%%%%%%%%%%%%%%
%%============================================================================%%
\section{INTRODUCTION}
% info10110349, villa2020survey,
Unmanned aerial vehicles (UAVs) have been employed in many application fields, including agriculture~\cite{hatanaka_agriculture}, transportation~\cite{Sekiguchi202190}, and search and rescue~\cite{Erdelj2017_SAR}. In these application domains, various types of UAVs are utilized, and most of the traditional ones can be categorized into fixed-wing or multi-rotor UAVs. Both of them have their own advantages. For example, a fixed-wing UAV tends to have a superior cruising distance, and a multi-rotor UAV can achieve static hovering, as discussed in~\cite{Boon17}. Hence, choosing an appropriate category of UAV for a task is a crucial factor for mission performance. 
Still, several applications require favorable properties of both fixed-wing and multi-rotor UAVs, e.g., transportation tasks tend to favor UAVs with long cruising distance, which only requires a small takeoff and landing area.

%Still, several applications require both favorable properties a fixed-wing and multi-rotor UAV has, e.g., a transportation task tends to favor a UAV with a long cruising distance, which only requires a small takeoff and landing area. 

%\red{上記の流れを解決する動きとして，Hybrid UAVが提案されつつある}

One of the prospective approaches to integrating beneficial features of both fixed-wing and multi-rotor UAVs is to design a hybrid UAV, which has a wing and is capable of vertical takeoff and landing (VTOL)~\cite{saeed_survey_2018}.
Generally, hybrid UAVs are classified into two classes: one is a tail-sitter, and the other is a convertible UAV.
A tail-sitter takes off and lands vertically on its tail, while a convertible UAV keeps its fuselage in a horizontal direction in all the cruising, landing, and takeoff phases.
While many tail-sitter UAVs have shown successful results~\cite{Kohno2014,9051852,Tal2023_tail_sitter}, convertible UAVs are more prevailing than the tail-sitter so far, partially due to the simple mechanism, smooth transitions, and more robust stability in the hovering phase, as mentioned in~\cite{saeed_survey_2018}.
% 本当は，もう少しactuatorの数とか色々と話したかった

%A convertible UAV keeps its fuselage in a horizontal direction in all the cruising, landing, and takeoff phases, while a tail-sitter takes off and lands vertically on its tail.

% これは制御手法なので，第三段落で引用？
Because of the variety of convertible UAVs' structure designs and challenges in control, various control methodologies are proposed to achieve cruising, VTOL, and phase transitions. 
In~\cite{Kikumoto2022}, a control strategy of transition from cruising to hovering flight is designed based on maneuverability analysis for a dual propulsion UAV, which has both upward and horizontally directed rotors for generating thrust for hovering and cruise flight.
Other than dual convertible UAVs, several studies have presented control methods for convertible UAVs having tiltable rotors or a wing~\cite{LIU2017135}.
The work~\cite{7152383} has proposed a controller for a convertible UAV with two tiltable rotors based on the gain scheduling technique. 
They have also presented another control strategy based on the total energy control system design~\cite{hernandez-garcia_total_2013}.
Nonlinear control techniques are also applied to convertible UAVs, e.g., nonlinear dynamic inversion~\cite{6300840} and nonlinear model predictive control~\cite{9444145,9183353}.
The fault tolerant control for an actuator failure is also considered in~\cite{mousaei_design_2022}.
All the above papers, which consider a convertible UAV with tiltable rotors or wings, necessitate additional actuators, such as servo motors, for controlling the tilting angle of rotors or wings. Although this additional mechanism makes it easy to control these tilting angles, it leads to an increase in the weight and could make the maintenance cumbersome.
Chiappinelli et al. integrated a passive rotary joint into the tilting mechanism, which decouples the tilting control \cite{chiappinelli_passive}. However, this UAV still requires control surfaces like an aileron or elevator, increasing weight and complicating control.
%\red{できれば，最後の文章に引用できる論文があるとよい}

% 提案手法の説明
This paper presents a novel convertible UAV with tiltable rotors, which control its rotors' tilting angles by rotor thrusts and a passive joint without additional actuators. 
%More concretely, the proposed UAV comprises a link connecting four rotors and a fuselage together with two rotors at the tail, as shown in Fig.~\ref{fig:topview}. 
%More concretely, the proposed UAV comprises a link connecting four rotors and a fuselage by a passive joint, together with two fixed rotors at the tail, as shown in Fig.~\ref{fig:topview}. 
More concretely, the proposed UAV owns four rotors mounted on a link, which is connected to the fuselage by a passive joint, and two fixed rotors at the tail, as shown in Fig.~\ref{fig:topview}.
This unique structure enables us to control the tilting angles of four rotors in front by adjusting their rotor thrusts. 
Therefore, the design reduces the weight of the tilting mechanism. % and enhances flexibility in tilt angle control. \blue{abstractから持ってきましたが，flexibilityの部分が何を意味しているのかがよく分かりませんでした}
We first derive the equation of motion of this UAV and analyze its controllability. % during \blue{static cruise flight.} %\blue{どのフェーズのcontrollabilityかを追記お願いします}
% We first derive the equation of motion of this UAV and analyze its controllability \blue{in the neighborhood of the static cruise flight.}
%
The derived condition for controllability is then utilized to plan the angle of attack of the body and the tilting angle of the link having four rotors. Second, we propose a controller that utilizes a tiltable link with four rotors to control the UAV's speed while the UAV's angle of attack follows the value planned from the equilibrium conditions. 
%Second, the control strategy for hovering, transition, and cruising phases are developed, where \blue{提案手法の特徴や良い点を書きたいです}.
Finally, the effectiveness of the proposed control method is demonstrated via simulation study.

\subsection{Notation}
A scaler variable is denoted by lower-case or upper-case symbols as $a$ or $A$.
A vector is denoted by a bold lower-case as $\bm a$, while a matrix is denoted by a bold upper-case as $\bm A$.
The $n\times m$ zero matrix is denoted as $\bm O_{n\times m}$.
The $n\times n$ zero matrix and identity matrix are denoted as $\bm O_n$ and $\bm I_n$, respectively.
The zero vector is denoted as $\bm 0$, whose size is obvious in each situation.
The rotational matrix corresponding to the $x$-axis is denoted as $\bm R_x$, similarly for the $y$- and $z$-axes.
$\cos (\ast)$ and $\sin (\ast)$ will be sometimes denoted as $C_\ast$ and $S_\ast$ for notational simplicity.

\section{MODELING} \label{sec:model}

\begin{figure}[t]
    % \begin{minipage}{0.48\linewidth}
        \centering
        \includegraphics[width=0.6\linewidth]{"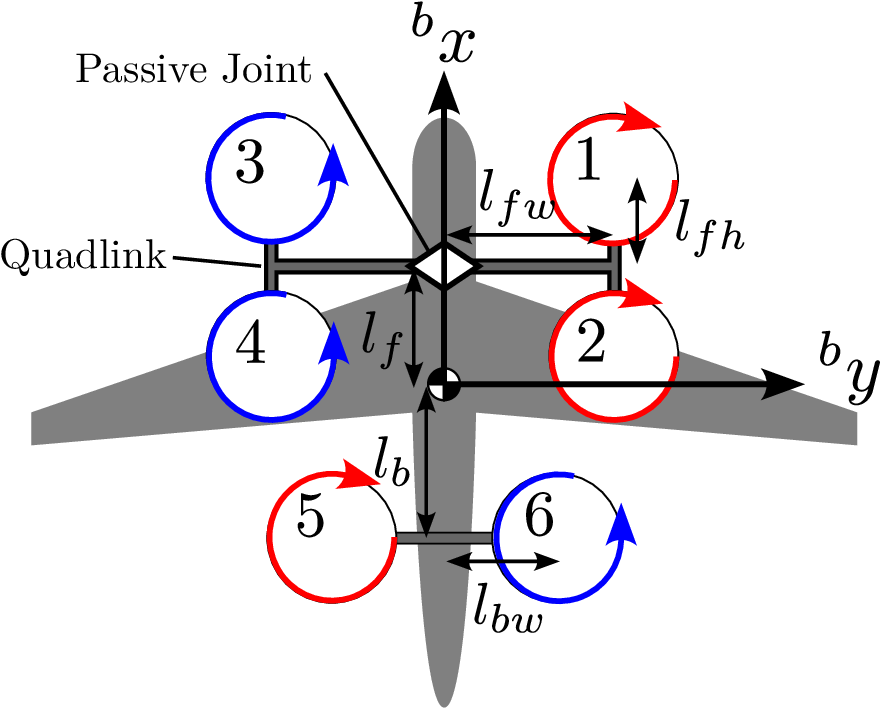"}
        \caption{Topview of the proposed UAV.}
        \label{fig:topview}
    % \end{minipage}\hspace{2em}
\end{figure}
\begin{figure}[t]
        % \begin{minipage}{0.48\linewidth}
        \centering
        \includegraphics[width=0.7\linewidth]{"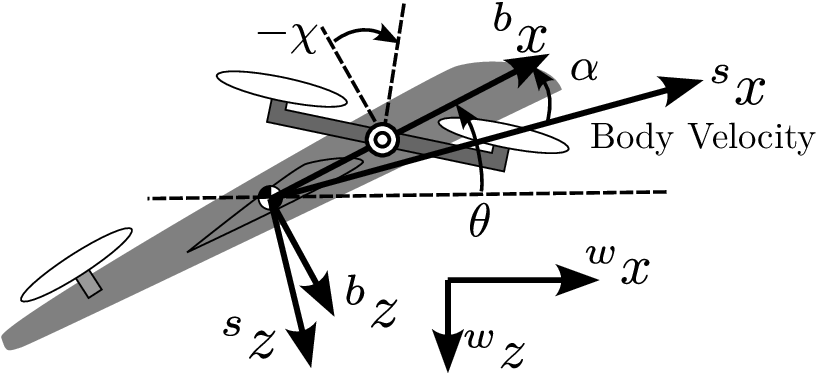"}
        \caption{Sideview of the proposed UAV.}
        \label{fig:sideview}
    % \end{minipage}   
\end{figure}
\begin{figure}[t]
        % \begin{minipage}{0.48\linewidth}
        \centering
        \includegraphics[width=\linewidth]{"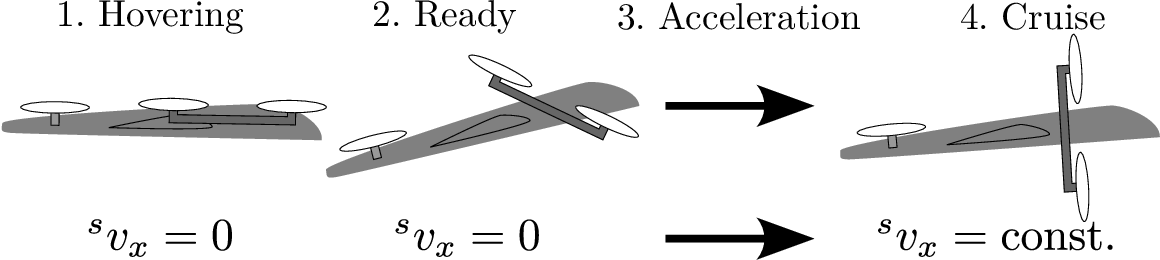"}
        \caption{Flight modes considered in this paper.}
        \label{fig:transition}
    % \end{minipage}   
\end{figure}

This section presents a novel design of a VTOL aerial vehicle shown in Fig.~\ref{fig:topview}. We first describe the overall picture of the proposed UAV while introducing several flight modes that will be considered in the paper. We then clarify that the dynamical equation of the UAV is composed of two factors: the rotor thrusts and the aerodynamics of the wing. The forces and torques rendered by these two factors are calculated in the later part of this section.

\subsection{Overview of the Vehicle} \label{ssec:mod_overview}

Let us first introduce several coordinate frames, which are illustrated in Fig.~\ref{fig:sideview}. 
The inertial frame $\Sigma_w$ is fixed to the ground aligned with the North-East-Down (NED) direction. The body-fixed frame $\Sigma_b$ is arranged at the center of mass of the vehicle so that its $x$-axis, $y$-axis, and $z$-axis are directed forward, right, and downward of the fuselage. 
In addition to these two frames, we introduce the stability frame~$\Sigma_s$, which is obtained by rotating the body frame $\Sigma_b$ around the $y$-axis to make the velocity in the $z$-axis of $\Sigma_s$ zero. Note that this rotational angle around the $y$-axis from $\Sigma_s$ to $\Sigma_b$ is referred to as the angle of attack, denoted as $\alpha$ hereafter.
The link frame $\Sigma_\ell$ is fixed to the center of the quadlink, and its rotation angle around the $y$-axis to the body frame is denoted as $\chi$ hereafter.
Each rotor also has its own coordinate frame, where its origin is located at the center of the rotor. The attitude of the first to fourth rotor frames aligns with the quadlink frame and the fifth and sixth with the body frame. 
%Each rotor also has its own coordinate frame, where its origin is located at the center of the rotor, and its attitude is the same as the quadlink.
%Each rotor can also define its own coordinate frame $\Sigma_{r_i}$ with the $z$-axis in the direction of the rotation axis, and the other orthogonal axis directed arbitrarily.
%
% \red{link coordinate$\Sigma_l$, $i$th rotor coordinate $\Sigma_{r_i}$を定義}
A left-side superscript of the variables signifies coordinate frame, where $b$, $w$, $s$, $\ell$, and $r_i$ mean the body, inertial, stability, link, and the $i$th rotor's frame, respectively.

The overview of the proposed VTOL UAV is depicted in Figs.~\ref{fig:topview} and \ref{fig:sideview}. 
The UAV comprises a fuselage, six rotors, and a link connecting four rotors and a fuselage by a passive joint. The link can freely rotate around the $y$-axis of $\Sigma_\ell$ and is referred to as the quadlink since it owns the first to fourth rotors. The remaining two rotors, fifth and sixth, are fixed to the tail of the body. 
The size of the quadlink and the rotors' configurations are defined with $l_f$, $l_{fw}$, $l_{fh}$, $l_b$, and $l_{bw}$, as shown in Fig.~\ref{fig:topview}.
Note that there are two types of rotors: a positive (P) rotor and a negative (N) rotor, depicted as red and blue in Fig.~\ref{fig:topview}. The P rotor rotates clockwise, and the N rotor counterclockwise.
The tilt angle of the quadlink with respect to the body frame is denoted as $\chi$. 
Its value reaches $0$ when the rotors are directed upward of the body and reaches $-\pi/2$ when they are oriented fully forward of the body. The control of $\chi$ is achieved by manipulating the thrust of the rotors within the quadlink.
%\red{PとNロータの配置について一言}

The proposed UAV distinguishes itself from conventional tiltrotor VTOL UAVs because of its unique tilting mechanism of rotors, which relies on a passive joint controlled solely by rotor thrusts. It is noteworthy that the presented UAV is controlled without any control surfaces, such as a rudder, aileron, or elevator. This design strategy allows us to eliminate additional actuators other than rotors, resulting in a reduction in overall weight and simplicity of structure.

%This vehicle distinguishes itself from conventional tiltrotor vehicles due to its unique tilt mechanism, which relies on a passive joint controlled solely by rotor thrusts. Notably, this vehicle is controlled without any control surfaces, such as a rudder, aileron, or elevator. This means that we eliminate the requirement for additional actuators, resulting in a reduction in overall weight.
% この機体は，conventionalなtiltrotorと違い，ロータ傾斜機構にpassive jointを採用してその傾斜角度の制御をロータの推力で行うところがuniqueである．さらに本論文ではエルロンやラダーなどのcontrol surfaceを導入せずに制御することに成功している．
% これらから，本機体は追加のアクチュエータなどが必要なく，重量増加を抑えることができる．

One of the most challenging aspects of VTOL UAV control lies in the transition from hovering mode to cruise flight mode. To address this challenge, we employ a four-phase transition strategy as shown in Fig.~\ref{fig:transition}. % and utilize gain scheduling techniques.
The four transition phases are as follows:
\begin{enumerate}
    \item Hovering Phase: During this phase, the vehicle takes off vertically and remains in a static hover in mid-air. The thrust of four rotors connected to quadlink is controlled to maintain $\chi=0$ during this phase.
    \item Ready Phase: The vehicle smoothly transitions to the planned initial $\alpha$ and $\chi$ suitable for starting the acceleration phase while the position of the UAV still keeps the same value as the hovering phase.
    \item Acceleration Phase: The vehicle accelerates forward during this phase until it reaches the desired forward velocity. For this goal, we design a control framework that tilts the quadlink forward to increase the UAV's speed while retaining the angle of attack to the desired value. We achieve this strategy by applying exact linearization, which eliminates the interference between the angle of attack and other state variables.
    %The vehicle accelerates forward during this phase until it reaches the desired forward velocity. Nonlinear aerodynamic effects change in response to the evolving vehicle states during the transition, making it challenging to control the vehicle with a single static controller. Therefore, we implement gain scheduling to adapt to these dynamic variations.
    \item Cruise Phase: The vehicle maintains stable flight while keeping the forward velocity constant value.
\end{enumerate}
% At first, the vehicle hovers at one point, then second, it transitions into the initial pose of third phase acceleration, finally settling on the cruise with a certain velocity. 
%The controller to achieve 
%Detailed controllers are described in Section 4.
The controller realizing these phases and transitions will be presented in Section~\ref{sec:controller}.

% この機体は，最終的にFig.~\ref{fig:transition}に示すように複数のフェーズを経て制御される．はじめに空中でホバリングしてから，加速の準備状態へと移行し，加速した後一定速度で飛行するcruise状態で落ち着く．それぞれのフェーズの詳細は4章のCONTROLLERで詳細に述べている．なお，これらのTransitionはすべてロータの推力のみで制御される，

% 本研究で扱う機体の概要をfig.\ref{fig:topview}, \ref{fig:sideview}に示す．
% 機体軸系には，機体の前方に$x$軸を，右側に$y$軸を，下方向に$z$をとる右手座標系を用いる．
% また，機体軸系を$y$軸回りに回転させ$z$軸方向の速度をゼロにした座標系を安定軸系という．
% このときの機体軸系と安定軸系の間の角度を迎角といい，$\alpha$で表す．

% 機体はロータ6枚で構成され，ロータのインデックスは図に示すとおりである．
% また，fig.\ref{fig:topview}の$l_{\ast}$はそれぞれ長さを示しており，飛行中に変わらない一定の値である．
% 機体前方の4枚はy軸方向の一軸周りに自由に回転するジョイントで接続されたリンク上に配置され，これをクアッドリンクと呼ぶ．
% クアッドリンクの機体に対する傾き角はロータの回転数を調整することで制御できる．
% この傾斜角を$\chi$で表す．なおロータの推力が機体上向き方向に向いているときを$\chi=0$とし，
% 完全に前方を向いた時に$\chi=-\pi/2$を取る．

% 翼の揚力と抗力が作用する中心点は機体重心と一致しており，空力による重心周りのモーメントは無視することができるとする．

\subsection{Equation of Motion}

The position of the vehicle in the inertial frame is denoted as ${}^{w}\bm x\in\mbr^3$. The orientation of the body frame in the inertial frame is represented with the rotation matrix $\bm R(\bm\eta)=\bm R_z(\psi)\bm R_y(\theta)\bm R_x(\phi)$ with the vector $\bm \eta = [\phi~\theta~\psi]^\top\in\mbr^3$, which combines the roll, pitch, and yaw angles of the vehicle.  
Let us also introduce the body velocity ${}^b\bm v\in\mbr^3$ and body angular velocity ${}^b\bm \omega \in\mbr^3$: the linear and angular velocity of the origin of the body frame relative to the inertial frame, as viewed in the current body fixed frame.

%Let us denote the position, the velocity, and the angular velocity of the vehicle in $\Sigma_w$ as $\bm x\in\mbr^3$, $\bm v\in\mbr^3$, and $\bm \omega \in\mbr^3$, respectively. The orientation of the body in $\Sigma_w$ is represented with the rotation matrix $\bm R(\eta)$ with the vector $\bm \eta = [\phi~\theta~\psi]\in\mbr^3$, which combines the roll, pitch, and yaw angles of the vehicle.  

%\blue{元の文章では，以下のように速度や角速度が in body frameで定義されると書いてありましたが，inertial frameでみた速度・角速度と書き直しても大丈夫ですか？body frameは機体と一緒に動くので，velocityやangular velocityをin the body frameで定義するのはおかしい気がしました．}
%\red{Let $\bm x\in\mbr^3$ represent the position of the vehicle in the inertial frame, $\bm v\in\mbr^3$ the velocity in the body frame, $\bm \eta = [\phi~\theta~\psi]\in\mbr^3$ the roll, pitch, and yaw angles, and $\bm \omega \in\mbr^3$ the angular velocity in the body frame.} 

As previously mentioned, the proposed vehicle has both rotors and a wing, and each of them generates the force and torque, whose point of application can be modeled to be the vehicle's center of mass. 
Let us denote the force and torque generated by all the rotors in $\Sigma_b$ as ${}^{b}\bm f_\text{rot}\in\mbr^3$ and ${}^{b} \bm \tau_\text{rot}\in\mbr^3$, which can be calculated by coordinate transformation and the summation of all rotors' thrusts, as detailed later. Also, we introduce the force and torque generated by the aerodynamic force of a wing in $\Sigma_s$ as ${}^{s}\bm f_\text{aero}\in\mbr^3$ and ${}^{s}\bm \tau_\text{aero}\in\mbr^3$, respectively. Then, the total thrust and moment around the center of mass of the vehicle can be expressed as
%
%Derived from Eq.~\eqref{eq:Mf_quad}, \eqref{eq:Mtau_quad}, \eqref{eq:faero}, and \eqref{eq:tauaero}, the total thrust and moment around the center of gravity of the vehicle can be expressed as follows.
\begin{align} 
    {}^{b} \bm f =& {}^{b} \bm f_\text{rot} + \bm R_y(\alpha)^\top {}^{s}\bm f_\text{aero} + \bm R(\bm\eta)^\top m\bm g, \label{eq:FB} \\
    {}^{b} \bm \tau =& {}^{b} \bm \tau_\text{rot} + \bm R_y(\alpha)^\top {}^{s} \bm \tau_\text{aero}, \label{eq:MB}
\end{align}
where $\bm g=[0~ 0~ g]^\top$ is the gravitational acceleration. 

The mass and the inertial tensor of the vehicle are denoted as $m\in\mbr$ and $\bm J\in\mbr^{3\times 3}$, respectively.
Then, the equation of motion for the vehicle can be expressed as 
\begin{subequations} \label{eq:equation_of_motioin}
\begin{align}
    & {}^{w}\dot{\bm x} = \bm R(\bm\eta) {}^{b}\bm v, \label{eq:navigation}\\
    & \dot{\bm\eta} = \begin{bmatrix}
        1 & \sin(\phi)\tan(\theta) & \cos(\phi)\tan(\theta) \\
        0 & \cos(\phi) & -\sin(\phi) \\
        0 & \sin(\phi)/\cos(\theta) & \cos(\phi)/\cos(\theta)
    \end{bmatrix} {}^{b}\bm \omega, \label{eq:kinematics}\\
    & m {}^{b}\dot{\bm v} = -{}^{b}\bm\omega\times (m {}^{b}\bm v) + {}^{b}\bm f, \label{eq:force}\\
    & \bm J {}^{b}\dot{\bm \omega} = -{}^{b}\bm\omega\times \bm J {}^{b}\bm \omega + {}^{b}\bm \tau  \label{eq:moment}.
\end{align}
\end{subequations}
In the following subsections, we derive ${}^{b}\bm f_\text{rot}$, ${}^{b} \bm \tau_\text{rot}$, ${}^{s}\bm f_\text{aero}$, and ${}^{s}\bm \tau_\text{aero}$ in Eqs.~\eqref{eq:FB} and \eqref{eq:MB}.

\subsection{Transformation of the Rotor Thrusts}

Each rotor generates the thrust force $f_i$ to the direction of the rotation axis while simultaneously rendering the counter torque $\tau_i$ to the opposite direction of the rotation. As in \cite{9393789}, we suppose that the proportional relationship holds between $\tau_i$ and $f_i$ as $\tau_i = \kappa_i f_i$, where $\kappa_i$ is named as the counter torque constant with $\kappa_i = \kappa$ for the P rotor and $\kappa_i = -\kappa$ for the N rotor. Let us define the wrench of $i$th rotor, which combines the thrust force and torque generated by $i$th rotor, seen in $\Sigma_{r_i}$, as 
\begin{align}
    {}^{r_i}{\bm f_i} = \begin{bmatrix}
        0 & 0 & f_i & 0 & 0 & \kappa_i f_i
    \end{bmatrix}^\top.
\end{align}
The force and torque generated by all rotors constituting the quadlink can be calculated by transforming the above wrench to the quadlink frame $\Sigma_{\ell}$ by employing the adjoint transformation~\cite{Hatanaka2015_Springer} and summing up all four rotors as 
\begin{align}
    \begin{bmatrix}
    {}^{\ell}\bm  f_\text{rot} \\ {}^{\ell}\bm \tau_\text{rot}
    \end{bmatrix} =
    \sum^4_{i=1} \begin{bmatrix}
        \bm I_3 & \bm O_3 \\
        \hat{\bm p}_{\ell r_i} & \bm I_3
    \end{bmatrix} {}^{r_i}\bm f_i,
\end{align}
where ${\bm p}_{\ell r_i}$ represents the relative position of $\Sigma_{r_i}$ with respect to $\Sigma_{\ell}$, and hereafter, the vector $\bm p_{\ast\ast}$ means the same way.
%\red{
%where ${\bm p}_{\ell r_i}$ represents the vector from the origin of $\Sigma_{\ell}$ to the origin of $\Sigma_{r_i}$ as viewed in $\Sigma_\ell$, and hareafter, the vector $\bm p_{\ast\ast}$ means the same way.
%}
Note that the operator $\hat{\ast}: \mbr^3 \to so(3):= \{ \bm S \in \mbr^{3\times 3} \mid\bm S + \bm S^\top = \bm O_3 \}$ provides $\hat{\bm a} \bm b = \bm a \times \bm b$ for any 3D vectors.

\begin{comment}
\violet{ロータは回転軸方向に推力$f_i$を生成し，同時に推力に比例したトルク$\kappa_i=\pm \kappa f_i$を回転方向の反対方向に生成するとする．
このときの$\kappa_i=\pm \kappa$は反トルク係数と呼ばれる比例係数であり，符号はCWのとき正，CCWのとき負である．
推力とトルクを合わせた6次元の一般化力の表現をレンチといい，このときロータ$i$座標系からみたロータ$i$のレンチは次のように表せる．
\begin{align}
    \bm f_i = \begin{bmatrix}
        0 & 0 & f_i & 0 & 0 & \pm\kappa_i f_i
    \end{bmatrix}
\end{align}
このレンチをAdjoint transformation\cite{murray1994mathematical}と呼ばれる変換行列を用いて，リンク座標系におけるレンチに変換し，quadlinkを構成するロータについて足し合わせると．
\begin{align}
    \begin{bmatrix}
    {}^{l}\bm  f_\text{rot} \\ {}^{l}\bm \tau_\text{rot}
    \end{bmatrix} =
    \sum^4_{i=1} \begin{bmatrix}
        \bm I_3 & \bm O_3 \\
        \hat{\bm p}_{link2rotor} & \bm I_3
    \end{bmatrix} \bm f_i
\end{align}
となる，ここで$\hat{\ast}$は歪対象行列を作る演算子である．}

リンクは機体に対してパッシブジョイントを介して回転する．この回転の摩擦を0と仮定すると，リンク座標のy軸周りのトルクはリンク傾斜角の制御のみに使われるとみなせる．
よって，$\chi$に働くトルクは次式で表せる．
\end{comment}

The quadlink is connected to the fuselage via a passive joint. Assuming the friction of this passive joint can be negligible, the torque around the $y$-axis of the quadlink is utilized only for the control of tilting angles of the quadlink. With this setting, the torque rotating the quadlink can be expressed as 
\begin{align}
    \tau_\chi = {}^{\ell}\tau_{\text{rot}y}. \label{eq:tau_chi}
\end{align}
The forces and the remaining torques are transmitted to the fuselage, which can be expressed in the body frame as 
\begin{align}
    \begin{bmatrix}
    {}^{b}\bm  f_\text{rot} \\ {}^{b}\bm \tau_\text{rot}
    \end{bmatrix} 
    =& \begin{bmatrix}
        \bm R_y(\chi) & \bm O_3 \\
        \hat{\bm p}_{b\ell}\bm R_y(\chi) & \bm R_y(\chi)
    \end{bmatrix}
    \begin{bmatrix}
    {}^{\ell}\bm  f_\text{rot} \\ {}^{\ell}\tau_{\text{rot}x} \\ 0 \\ {}^{\ell}\tau_{\text{rot}z}
    % \bm  f_\text{link} \\ \tau_{\text{link}x} \\ 0 \\ \tau_{\text{link}z}
    \end{bmatrix}\notag\\
    & + \sum^6_{i=5} \begin{bmatrix}
        \bm I_3 & \bm O_3 \\
        \hat{\bm p}_{b r_i} & \bm I_3
    \end{bmatrix} {}^{r_i} \bm f_i.\label{eq:bodywrench}
\end{align}

Let us define the rotor thrust vector as $\bm f_\text{rot} = [f_1 \cdots f_6]^\top$. Then, Eqs.~\eqref{eq:tau_chi} and \eqref{eq:bodywrench} can be expressed in a matrix form as
% \violet{
\begin{align}
    \begin{bmatrix}
        {}^{b} \bm f_\text{rot} \\ {}^{b} \bm \tau_\text{rot}
    \end{bmatrix}
    =&
    \begin{bmatrix}
        \bm \lambda_{1} & \bm \lambda_{1} & \bm \lambda_{2} & \bm \lambda_{2} & \bm \lambda_{3} & \bm \lambda_{4}
    \end{bmatrix} 
    \bm f_\text{rot}, \label{eq:Mf} \\
    \tau_\chi =&
    \begin{bmatrix}
        l_{fh} & -l_{fh} & l_{fh} & -l_{fh}
    \end{bmatrix}
    \bm f_\text{rot}, \label{eq:Mchi}
\end{align}
where 
\begin{align}
    \bm \lambda_{1} \!\!=\!& \begin{bmatrix}
    -S_\chi\! & \!\!0\! & \!\!-C_\chi\! & \!\!-l_{fw} C_\chi\!-\!\kappa S_\chi\! & \!\!l_{f} C_\chi\! & \!\!l_{fw} S_\chi\!-\!\kappa C_\chi
    %    - S_\chi & 0 & - C_\chi & -l_{fw} C_\chi\!-\!\kappa S_\chi & l_{f} C_\chi & l_{fw} S_\chi\!-\!\kappa C_\chi
    \end{bmatrix}^\top,\notag\\
    \bm \lambda_{2} \!\!=\!& \begin{bmatrix}
    - S_\chi\! & \!\!0\! & \!\!-C_\chi\! & \!\!l_{fw} C_\chi\!+\!\kappa S_\chi\! & \!\!l_{f} C_\chi\! & \!\!-l_{fw} S_\chi\!+\!\kappa C_\chi
    %    - S_\chi & 0 & - C_\chi & l_{fw} C_\chi\!+\!\kappa S_\chi & l_{f} C_\chi & -l_{fw} S_\chi\!+\!\kappa C_\chi
    \end{bmatrix}^\top,\notag\\
    \bm \lambda_{3} \!\!=\!& \begin{bmatrix}
        0 & 0 & -1 & -l_{bw} & -l_b & \kappa
    \end{bmatrix}^\top,\notag\\
    \bm \lambda_{4} \!\!=\!& \begin{bmatrix}
        0 & 0 & -1 & l_{bw} & -l_b & -\kappa
    \end{bmatrix}^\top.\notag
\end{align}
Notice that the first and second columns in \eqref{eq:Mf} are the same. Similarly, the third and fourth columns in \eqref{eq:Mf} are equal. This signifies that the wrench of two rotors on the same side of the quadlink is equivalent. This fact motivates us to regard the two rotors on the same side as one P or N rotor by introducing the rotor input vector $\bm f_\text{r+}= [f_1\!+\!f_2~f_3\!+\!f_4~f_5~f_6]^\top$. Then, the UAV can be modeled like a quad-rotor UAV with two tiltable rotors in front as 
%
%\eqref{eq:Mf}の1と2行目，3と4行目がそれぞれ同じベクトルになっている．
%これは，ふたつのロータの推力から機体にもたらされるレンチが等しいことを表しており，推力の和を考えることで一つのロータであるかのように扱うことができる．
%これら二組のロータの和を使って新しいロータ入力ベクトルを$\bm f_\text{r+}= [f_1\!+\!f_2,~f_3\!+\!f_4,~f_5,~f_6]^\top$とすると，機体レンチをの式\eqref{eq:bodywrench}はクアッドコプターのように扱うことができる．
%さらに，$\bm\lambda_i,~i\in\{1,\cdots,4\}$の2行目がすべて0になっているため，機体のy軸方向の推力は生成しないことも考慮するとEq.~\eqref{eq:Mf}は次のように書き換えることができる．
% We define new rotor input vectors as $\bm f_\text{r+}= [f_1+f_2,~f_3+f_4,~f_5,~f_6]^\top,\bm f_\text{r-}= [f_1-f_2,~f_3-f_4]^\top$. Consequently, Eq.~\eqref{eq:Mf}, \eqref{eq:Mtau}, and \eqref{eq:Mchi} can be rewritten as follows.
\begin{align}
    &\begin{bmatrix}
        {}^{b} f_{\text{rot}x} \\
        {}^{b} f_{\text{rot}z} \\
        {}^{b} \bm \tau_\text{rot}
    \end{bmatrix}
    = \bm M(\chi) \bm f_\text{r+}, \label{eq:M_quad} \\ 
    &\bm M 
    = \begin{bmatrix}
        - S_\chi & - S_\chi & 0 & 0 \\
        - C_\chi & - C_\chi & -1 & -1 \\
        -l_{fw} C_\chi-\kappa S_\chi &l_{fw} C_\chi+\kappa S_\chi & -l_{bw} & l_{bw} \\
        l_{f} C_\chi & l_{f} C_\chi & -l_b & -l_b \\
        l_{fw} S_\chi-\kappa C_\chi & -l_{fw} S_\chi+\kappa C_\chi & \kappa & -\kappa
    \end{bmatrix}. \notag
    % \bm M(\chi)=& 
    % \begin{bmatrix}
    %     \bm \lambda_{1} & \bm \lambda_{2} & \bm \lambda_{3} & \bm \lambda_{4}
    % \end{bmatrix},\notag
\end{align}
Note that we have eliminated the second row of \eqref{eq:Mf}, which represents the force in $y$-direction, since it always results in zero because of the second row of $\bm\lambda_i,~i\in\{1,\cdots,4\}$ being zero.
%Note that we eliminate the second row of \eqref{eq:Mf}, corresponding to the force in $y$-direction because it always becomes zero as the second row of $\bm\lambda_i,~i\in\{1,\cdots,4\}$ are zero.
%
Let us also introduce the thrust input vector $\bm f_\text{r-}= [f_1\!-\!f_2~f_3\!-\!f_4]^\top$. Then, Eq.~\eqref{eq:tau_chi} can be simplified as 
%他方，ロータの推力の差を使って$\bm f_\text{r-}= [f_1\!-\!f_2,~f_3\!-\!f_4]^\top$を定義すると，リンクトルクの式\eqref{eq:tau_chi}を単純化して考えることができる．
\begin{align}
    \tau_\chi =& 
    \begin{bmatrix}
        l_{fh} & l_{fh}
    \end{bmatrix}
    \bm f_\text{r-}. \label{eq:Mchi_quad}    
\end{align}
% }
Given a desired wrench $[{}^{b}f_{\text{rot}x}~{}^{b}f_{\text{rot}z}~{}^{b}\bm\tau_\text{rot}^\top]^\top$ and the torque $\tau_\chi$ for the quadlink, the rotor thrust realizing them can be derived by the inverse transformation of Eqs.~\eqref{eq:M_quad} and \eqref{eq:Mchi_quad}. 

\begin{comment}
\violet{
コントローラにより目的の機体レンチ$[{}^{b}f_{\text{rot}x}~{}^{b}f_{\text{rot}z}~{}^{b}\bm\tau_\text{rot}^\top]^\top$とリンクトルク$\tau_\chi$が与えられたときに，Eq. \eqref{eq:M_quad} and \eqref{eq:Mchi_quad}を使ってそれぞれを実現するロータ推力を求めることができる．
なお，Eq. \eqref{eq:M_quad}の行列の次元は$\mbr^5\times\mbr^4$であるが，$\chi$を動かすことで$[{}^{b}f_{\text{rot}x}~{}^{b}f_{\text{rot}z}~{}^{b}\bm\tau_\text{rot}^\top]^\top\in\mbr^5\to(\chi,~\bm f_\text{r+})\in\mbr^5$のように厳密な逆変換が存在する．
% Once the desired wrench for the vehicle and the torque for the quadlink are determined, rotor thrusts that generate these desired values can be calculated by inversely transforming Eq. \eqref{eq:M_quad} and \eqref{eq:Mchi_quad}. 
% It's important to note that the rank of the matrix in Eq. \eqref{eq:M_quad} is 4, meaning that an inverse transformation may not always exist. However, by adjusting the value of $\chi$, we can derive an equation that allows for a strict inverse transformation.
}    
\end{comment}

%\red{
\begin{remark} \label{rem:quadlink}
    Because $\bm f_\text{r+}$ and $\bm f_\text{r-}$ have six elements with six rotors' thrusts, we can find one combination of rotors' thrusts for any pair of $\bm f_\text{r+}$ and $\bm f_\text{r-}$. Also, by definitions, any elements in $\bm f_\text{r+}$ and $\bm f_\text{r-}$ do not depend on other elements.
    %$\bm f_\text{r+}$と$\bm f_\text{r-}$は定義より，一方のどの要素の値も他方の値に依存することはない．
    %また，$(\bm f_\text{r+},~\bm f_\text{r-})\leftrightarrow\bm f_\text{rot}$の変換は一意（全単射）である．
    %Eq. \eqref{eq:M_quad} and \eqref{eq:Mchi_quad}から$\bm f_\text{r+}$と$\bm f_\text{r-}$を別々に求めたとき，それらに対応する$\bm f_\text{rot}$を一意に求めることができる．
    % The control of the rigid body vehicle and the tilt angle of the quadlink are separable. From Eq.~\eqref{eq:Mchi_quad}, the thrust of the 5th and 6th rotors in $\bm f_\text{r-}$ can have arbitrary values. Therefore, a rotor thrusts $\bm f_\text{rot}$ corresponding to any $\bm f_\text{r+}$ and $\bm f_\text{r-}$ exists.
\end{remark}

\subsection{Aerodynamic Force}
As the velocity of the vehicle increases, aerodynamic forces serve as an important factor. 
Suppose that the point of application of the aerodynamic force is the same as the vehicle's center of mass. 
Then, the aerodynamic force and torque in $\Sigma_s$ are defined as
\begin{align}
    {}^{s} \bm f_\text{aero} =& \begin{bmatrix}
                                  -D & 0 & -L
                              \end{bmatrix}^\top, \label{eq:faero} \\
    {}^{s} \bm \tau_\text{aero} =& \bm 0, \label{eq:tauaero}
\end{align}
with 
\begin{align}
    D =& \frac{1}{2} \rho {}^{s}v_x^2 \mathcal{S} \mathcal{C}_D, \label{eq:D}\\
    L =& \frac{1}{2} \rho {}^{s}v_x^2 \mathcal{S} \mathcal{C}_L, \label{eq:L}
\end{align}
that represent the drag and lift forces, respectively.
Note that $\rho$ represents air density, $\mathcal{S}$ is the wing area, and $\mathcal{C}_D$ and $\mathcal{C}_L$ are coefficients dependent on the angle of attack $\alpha$.
To determine $\mathcal{C}_D$ and $\mathcal{C}_L$, we approximate the results obtained from the work \cite{ozdemir_design_2014} using polynomial expressions.

\section{REALIZABILITY OF THE CRUISE FLIGHT} \label{sec:cruisability}

\subsection{Realizability Analysis} \label{ssec:realize}

In this subsection, we analyze the feasibility of the cruise flight of the proposed VTOL UAV. We first define the cruise flight and clarify the conditions necessary for achieving it. Because the derived conditions provide appropriate values for the angle of attack and quadlink, $\alpha$ and $\chi$, the acquired conditions will be utilized to plan $\alpha$ and $\chi$ during the flight in the next subsection.

Let us define the cruise flight considered in this paper.
\begin{definition}[Cruise flight] \label{def:cruise}
    Given a cruise speed ${{}^{s}v_x\!>\!0}$, the UAV is regarded in the cruise flight if the following conditions are satisfied.
    \begin{align*}
        {}^w \dot{\bm x}=[{}^{s}v_x~0~0]^\top,~{}^{b}\dot{\bm v}=\bm 0,~\dot{\bm\eta} = \bm 0,~{}^b \dot{\bm\omega} =\bm 0.
    \end{align*}
\end{definition}
Note that the above definition implies that the UAV in the cruise phase flies at a constant altitude because its velocity is directed horizontally. In this setting, the $x$-$y$ plane of the inertial frame $\Sigma_w$ is parallel to the $x$-$y$ plane of the stability frame $\Sigma_s$, resulting in a complete alignment between the angle of attack $\alpha$ and the pitch angle $\theta$. Then, the equilibrium point in the cruise flight phase can be expressed as 
\begin{subequations} \label{eq:equil_point}
\begin{align} 
    {}^{b}\bm v =& \begin{bmatrix}
        {}^{s}v_x\cos{\alpha} & 0 & {}^{s}v_x\sin{\alpha}
    \end{bmatrix}^\top,\\
    \bm\eta =& \begin{bmatrix}
        0 & \alpha & 0
    \end{bmatrix}^\top,\\
    \bm \omega =& \begin{bmatrix}
        0 & 0 & 0
    \end{bmatrix}^\top.\label{eq:equilibrium}
\end{align}
\end{subequations}
% \red{上の１つ目の式，色々と間違っていませんか？}
% なお，${}^{s}v_x=0$のときがHoveringである．さらに，本論文では簡単のためクルーズ飛行は慣性座標系で水平に飛行している状態とする．すなわち，高度が一定で飛行しており，安定軸系と慣性座標のx-y平面が平衡であることを示している．このとき，迎角$\alpha$とピッチ角$\theta$は完全に一致する．

%The condition for achieving cruise flight can be derived from a similar procedure to the hoverability condition for multi-rotor UAVs, which is introduced as follows.

Let us denote the capability of the cruise flight, in the sense of Definition~\ref{def:cruise}, as \textit{cruisability} hereafter. Then, the cruisability of the proposed UAV can be analyzed by the following Lemma.

\begin{lemma}[Cruisability check]\label{lem:cruisability}
    The proposed UAV achieves a cruise flight if and only if the following two conditions are satisfied.
    \begin{enumerate}
        \item \label{cruise_cond1}
            There exists input vector $\bm f_{r+}$ satisfying
            \begin{align} 
                \bm M \bm f_{r+} = 
                \begin{bmatrix} 
                    -\bm R_y(\alpha)^\top( {}^{s}\bm f_\text{aero} + m\bm g)\\
                    \bm O_{3\times 1} 
                \end{bmatrix}.\label{eq:cond1}
            \end{align}
        \item \label{cruise_cond2}
        The control effectiveness matrix $\bm M$ satisfies
            \begin{align*} \mathrm{rank}(\bm M) = 4. \end{align*}
    \end{enumerate}
\end{lemma}
The first condition in Lemma~\ref{cruise_cond1} indicates the existence of the equilibrium points, namely the existence of the thrust inputs that cancel out the combined force of gravity and the aerodynamic force, as in Fig.~\ref{fig:sideview_force}. % and stay at a fixed point in mid-air. 
The second condition assures the local controllability around the equilibrium point. Please refer to the authors' antecessors' work~\cite{9393789}, which considers the hovering capability of the multi-rotor UAV with upward-oriented rotors, for details of the connection between the local controllability and the second condition.

For a given cruising speed, the above requirements provide us with the appropriate values of the angle of attack $\alpha$ and quadlink $\chi$ of the proposed UAV, as detailed in the next theorem.
%The above requirements provide us with the appropriate values of the angle of attack $\alpha$ and quadlink $\chi$ of the proposed UAV for a given cruising speed, as detailed in the next theorem.
%
% \red{The below subsections describe detailed analyses of these two conditions.
\begin{theorem} \label{thm:cruise}
    The proposed UAV realizes a cruise flight when the following conditions are satisfied. 
    \begin{enumerate}
        \item ${}^{s}v_x$, $\alpha$, and $\chi$ satisfy
            \begin{align}
                (mg-L)& \left(S_{\alpha+\chi}+\frac{l_f}{l_b}C_\chi S_\alpha\right) \notag\\
                + D& \left(C_{\alpha+\chi}+\frac{l_f}{l_b}C_\chi C_\alpha\right) = 0. \label{eq:alphachi_cond}
            \end{align}
        \item $\chi \neq \tan^{-1}(\kappa/l_{fw})$.
    \end{enumerate}
\end{theorem}
\begin{proof}
Each of the above two conditions corresponds with items~1 and~2 in Lemma~\ref{lem:cruisability}. 
In the following, we derive each condition in order.

First, let us transform the condition~\eqref{eq:cond1} defined in the inertial frame into the stability frame as 
    \begin{align}
        \begin{bmatrix}
            \bm R_y(\alpha) & \bm O_3 \\
            \bm O_3         & \bm I_3
        \end{bmatrix}\bm M \bm f_{r+} + 
        \begin{bmatrix}
            \bm {}^{s}\bm f_\text{aero} + m\bm g\\
            \bm O_{3\times 1} 
        \end{bmatrix}
        = \bm 0.
    \end{align}
    By substituting the equilibrium point~\eqref{eq:equil_point} and allocating the same magnitude of thrust input symmetrically with respect to the $x$-axis of the body frame, namely $f_1=f_3,~f_2=f_4,~f_5=f_6$, we obtain 
    %平衡点\eqref{eq:equilibrium}を代入し，対称の位置にあるロータに同じ推力，すなわち$f_1=f_3,~f_2=f_4,~f_5=f_6$を代入すると，
    % さらに，もともとy軸方向の推力は生成しないことから，平衡点入力は次の3式を満たせばよいことになる．
    \begin{align} \label{eq:alpha_chi_cond_bf}
        \begin{bmatrix}
            -D - f_f\sin(\alpha+\chi) - f_b\sin(\alpha) \\
            0 \\
            mg - L - f_f\cos(\alpha+\chi) - f_b\cos(\alpha) \\
            0 \\
            f_f l_f \cos(\chi) - f_b l_b \cos(\chi) = 0 \\
            0 
        \end{bmatrix} = 
        \begin{bmatrix}
            0 \\ 0 \\ 0 \\ 0 \\ 0 \\ 0
        \end{bmatrix},
    \end{align}
    where $f_f = \sum^4_{i=1}f_i$ and $f_b = \sum^6_{i=5}f_i$.
    Note that Eq.~\eqref{eq:alpha_chi_cond_bf} represents the condition where the force in the $x$ and $z$ direction and the torque around the $y$-axis are zero, as shown in Fig.~\ref{fig:sideview_force}.
    %これはx方向の力，z方向の力，y軸回りのモーメントが釣り合っている状態を表している．
    %\red{fig.\ref{fig:sideview_force}を参照する．}
    \begin{figure}[t]
        % \begin{minipage}{0.48\linewidth}
            \centering
            \includegraphics[width=0.6\linewidth]{"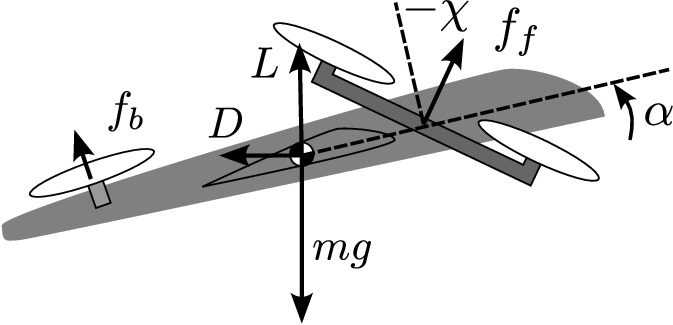"}
            \caption{The relationship of the forces at the equilibrium point.}
            \label{fig:sideview_force}
        % \end{minipage}\hspace{2em}
    \end{figure}
    From \eqref{eq:alpha_chi_cond_bf}, we can obtain the condition \eqref{eq:alphachi_cond} in the first item.
    % \begin{align}
    %     (mg-L) \left(S_{\alpha+\chi}+\frac{l_f}{l_b}C_\chi S_\alpha\right)
    %     + D \left(C_{\alpha+\chi}+\frac{l_f}{l_b}C_\chi C_\alpha\right) = 0. \label{eq:alphachi_cond}
    % \end{align}
   %Eq.~\eqref{eq:alphachi_cond}を満たす${}^{s}v_x\alpha,~\chi$の集合は，Fig.~\ref{fig:equilibrium}に示すような平面集合になる．
    %このとき，機体には平衡点入力が存在し，Lemma~\ref{lem:cruisability}(\ref{cruise_cond1})を満たす．
    \begin{figure}[t]
        % \begin{minipage}{0.48\linewidth}
            \centering
            \includegraphics[width=0.8\linewidth]{"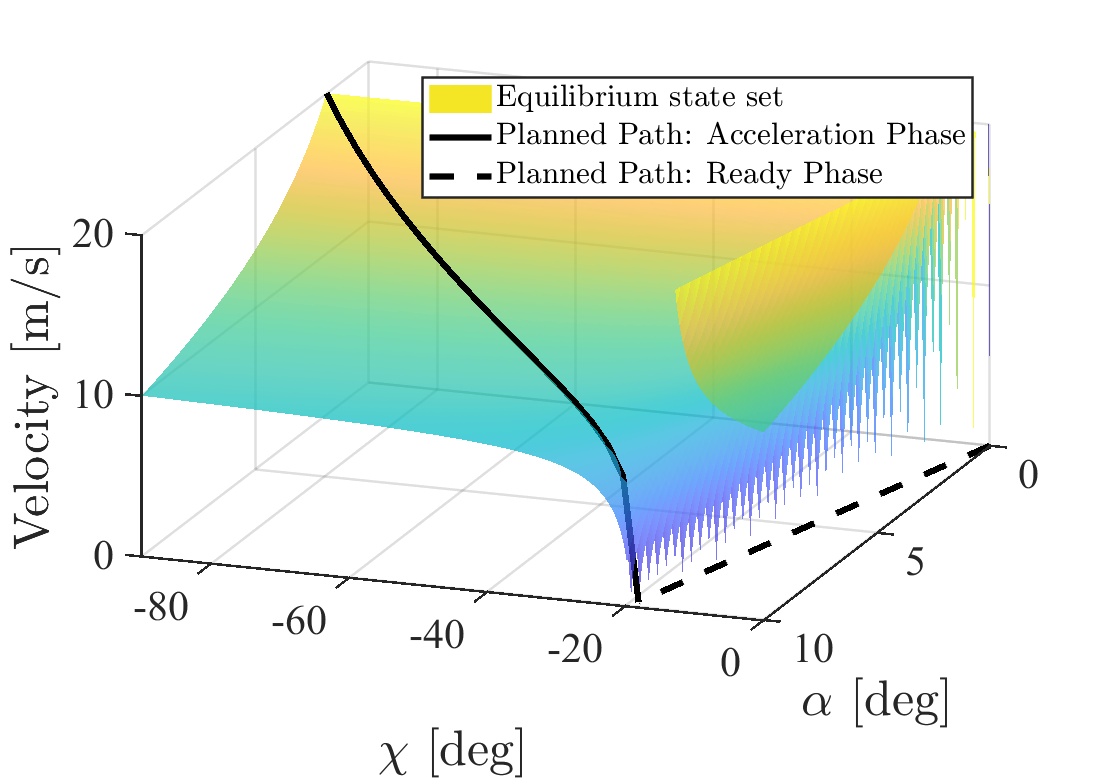"}
            \caption{The curve satisfying cruisability conditions.}
            \label{fig:equilibrium}
        % \end{minipage}\hspace{2em}
    \end{figure}

Next, let us derive the condition for satisfying item 2 in Lemma~\ref{lem:cruisability}. 
By splitting Eq.~\eqref{eq:M_quad} as
    \begin{align}
        \bm M 
        =& \left[\begin{array}{cc|cc}
            - S_\chi & - S_\chi & 0 & 0 \\
            \hline
            - C_\chi & - C_\chi & -1 & -1 \\
            -l_{fw} C_\chi-\kappa S_\chi &l_{fw} C_\chi+\kappa S_\chi & -l_{bw} & l_{bw} \\
            \hline
            l_{f} C_\chi & l_{f} C_\chi & -l_b & -l_b \\
            l_{fw} S_\chi-\kappa C_\chi & -l_{fw} S_\chi+\kappa C_\chi & \kappa & -\kappa
        \end{array}\right] \notag\\
        = & 
        \left[\begin{array}{c|c}
            \bm M_{11} & \bm M_{12} \\
            \hline
            \bm M_{21} & \bm M_{22} \\
            \hline
            \bm M_{31} & \bm M_{32}
        \end{array}\right],
    \end{align}
    the following condition holds.
    \begin{align}
        \mathrm{rank}(\bm M)&=
        \mathrm{rank}(\bm M_{22}) + \mathrm{rank}\left(\begin{bmatrix}\bm M_{11}\\ \bm M_{31}\end{bmatrix}\right).
    \end{align}
    Note that $\mathrm{rank}(\bm M_{22})=2$ from the definition. In addition, $\mathrm{rank}([\bm M_{11}^\top \bm M_{31}^\top]^\top)\!=\!2$ holds when $\chi \!\neq\! \tan^{-1}(\kappa/l_{fw})$ is satisfied, resulting in $\mathrm{rank}(\bm M)\!=\!4$.
    % $\chi\in [-\pi/2,~0]$ is satisfied, resulting in $\mathrm{rank}(\bm M)=4$.
This completes the proof.
 %   となり，自明に$\mathrm{rank}(\bm M_{22})=2$である．
 %   また，$\chi=[-\pi/2,~0]$のとき$\mathrm{rank}(\bm M_{31})=2$となり，$\mathrm{rank}(\bm M)=4$となる
 %   以上でLemma~\ref{lem:cruisability}の二つの条件を満たす$\alpha,~\chi,~{}^{s}v_x$についての条件を導出できた．
\end{proof}

Figure~\ref{fig:equilibrium} illustrates the curve satisfying both conditions in Theorem~\ref{thm:cruise} in a space with the $x$, $y$, and $z$ axes representing $\chi$, $\alpha$, and the cruising speed ${}^{s}v_x$.
%The curve shown in Fig.~\ref{fig:sideview_force} corresponds with the equ
On the curve, the equilibrium input exists, and the system can be locally stabilized. % by the linear state feedback controller.

% \subsection{平衡点入力の存在条件}
% 機体座標系と安定軸系間の回転$\bm R_y(\alpha)$を用いて
% eq.\ref{eq:navigation}-\ref{eq:moment}を書き換え，整理すると．
% \begin{align}
%     & \dot{\bm x} = \bm R(\eta) \bm R_y^\top(\alpha) {}^{s}\bm v, \label{eq:navigation_s}\\
%     & \dot{\bm\eta} = \begin{bmatrix}
%         1 & \sin(\phi)\tan(\theta) & \cos(\phi)tan(\theta) \\
%         0 & \cos(\phi) & -\sin(\phi) \\
%         0 & \sin(\phi)/\cos(\theta) & \cos(\phi)/\cos(\theta)
%     \end{bmatrix} \bm R_y^\top(\alpha) {}^{s}\bm \omega, \label{eq:kinematics_s}\\
%     & m{}^{s}\dot{\bm v} = -{}^{s}\bm\omega\times (m{}^{s}\bm v) + {}^{s}\bm f, \label{eq:force_s}\\
%     & \tilde{\bm J}{}^{s}\dot{\bm \omega} = -{}^{s}\bm\omega\times \tilde{\bm J} {}^{s}\bm \omega + {}^{s}\bm \tau  \label{eq:moment_s}.
% \end{align}
% となる．ここで，$\tilde{\bm J}=\bm R_y(\alpha)\bm J \bm R_y^\top(\alpha)$である．
% これらに平衡点${}^{s}\bm v=[{}^{s}v_x~0~0]^\top,~{}^{s}\bm\omega=\bm 0,~\eta=[0~\alpha~0]^\top$を代入すると，eq.\eqref{eq:navigation}, eq.\eqref{eq:kinematics}は恒等式になり，
% eq.\eqref{eq:force},~eq.\eqref{eq:moment}は次のようになる．
% \begin{align}
%     {}^{s}\bm f = \bm 0 \\
%     {}^{s}\bm \tau = \bm 0
% \end{align}

% \subsection{可制御性}
% \begin{theorem}[Realizability]
%     Eq.~\eqref{eq:alphachi_cond}を満たし，かつ$\chi=[-\pi/2,~0]$のとき，機体はクルーズ飛行可能である．
% \end{theorem}
% \begin{proof}

% \end{proof}

\subsection{Planning of $\alpha$ and $\chi$} \label{ssec:plan}

% \begin{figure}[t]
%     % \begin{minipage}{0.48\linewidth}
%         \centering
%         \includegraphics[width=\linewidth]{"figure/controllability/equilibrium_topview.eps"}
%         \caption{平行点入力が存在する$\alpha,~\chi$のマップ．}
%         \label{fig:equilibrium_topview}
%     % \end{minipage}\hspace{2em}
% \end{figure}

During the transition phase, the UAV is required to change its angle of attack $\alpha$ and tilt angle of quadlink $\chi$ to accelerate its body. However, $\alpha$ and $\chi$ have to take close enough value to the condition derived in Theorem~\ref{thm:cruise} to make the UAV controllable. To achieve this goal, we utilize the curve in Fig.~\ref{fig:equilibrium}, derived from Theorem~\ref{thm:cruise}, for planning $\alpha$ and $\chi$. In Fig.~\ref{fig:equilibrium}, the black line and dashed line represent the planned path for acceleration and ready phase, respectively. 
For both transitions, we design paths so that $\alpha$ and $\chi$ change linearly. %, as in Fig.~\ref{fig:equilibrium_topview}.

As mentioned at the end of Section~\ref{ssec:mod_overview} and Fig. 3, during the ready phase, the UAV tilts its quadlink forward to prepare for the acceleration phase. 
At the same time, the angle of attack has to increase to keep the cruising speed ${}^s v_{x}= 0$ during the ready phase and acquire enough aerodynamic force to maintain its altitude even with a small cruising speed at the beginning of the subsequent acceleration phase.
%At the same time, the angle of attack has to increase to keep the cruising speed ${}^s v_{x}= 0$ and acquire enough lift force at the beginning of the acceleration phase.
Once the UAV moves to the acceleration phase, the UAV increases its cruising speed by tilting quadlink forward more. As the cruising speed becomes large, a small angle of attack is enough to generate necessary lifting forces. 
%\red{During the acceleration phase, 下記のコメントでも書いた通り，Simulationで見せたいことを書きたい}

Note that the point $(\chi, \alpha, {}^s v_x)$ on the curve in Fig.~\ref{fig:equilibrium} signifies the equilibrium point at which the thrust input achieving the cruise flight exists with the designated $\chi$ and $\alpha$. 
Nevertheless, in the acceleration phase, the UAV is required to increase its cruise speed rather than stay at the equilibrium point. To achieve this goal, we have to shift the value of $\chi$ or $\alpha$ from the equilibrium point. 
In this paper, we opt for changing $\chi$ rather than $\alpha$ because the derivation in the angle of attack causes the change of aerodynamic force, which necessitates a more complicated analysis as Eqs.~\eqref{eq:D} and~\eqref{eq:L} are nonlinear. 
The controller in the following section is designed to control $\chi$ to accelerate the vehicle rather than changing $\alpha$ from its value at the equilibrium point.

\begin{comment}
上記の結果を用いて機体のtransition時の迎角$\alpha$，チルト角$\chi$を計画する．
Fig.~\ref{fig:equilibrium}の平面上にあるときにcruise flight可能であり，この平面上でtransitionの時系列変化をプランニングする．
この平面上で目的関数等を設定して最適パスを計画することも考えられるが，本論文の主旨は機体の提案とクルーズ可能条件の証明，シミュレーションによる決定のみであるため，最適設計は今後の課題とする．
本シミュレーションでは，$\alpha$と$\chi$平面への射影が線形に変化するパスを考える．
\red{詳しく説明or Fig5の上から見た図を添付（後者がわかりやすいかな）}

なお，これは平衡点に限った入力であり，動的に制御するにはこの値では不十分である．いずれかの変数を使って制御する必要がある．
${}^{s}v_x,~\alpha$はEq.~\eqref{eq:D},~\eqref{eq:L}に示す通り空力に非線形に関係する変数であり，値を変化させると空力が変化してダイナミクスが複雑になってしまう．そのため，シミュレーションでは$\chi$を変化させて制御する．詳細な説明はcontrollerの章にしめす．
\end{comment}

% \end{document}

\section{CONTROLLER} \label{sec:controller}

% \subsection{Control Strategy}
\begin{figure*}[t]
        \centering
        \includegraphics[width=\linewidth]{"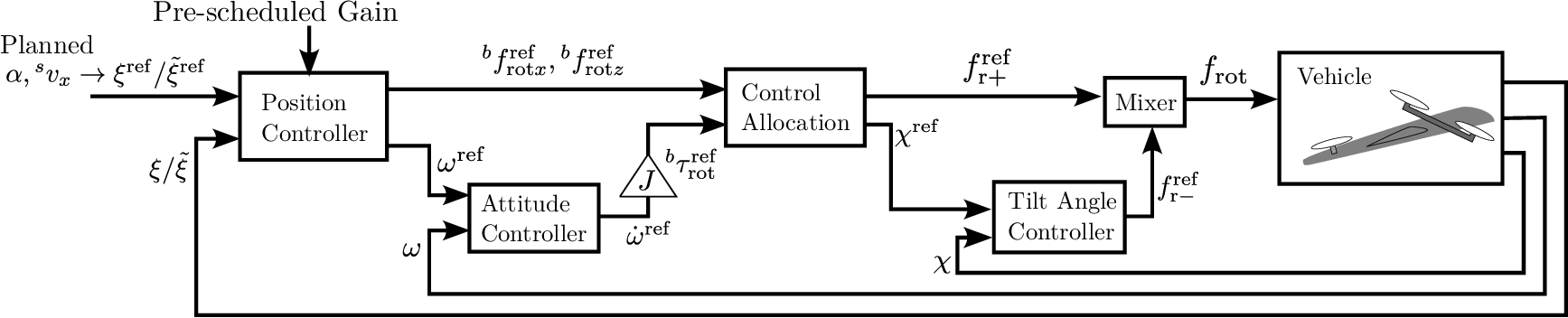"}
        \caption{The overall architecture of the proposed controller.}% \red{control allocationの違い，LQRが上位のplanningで決まることをわかりやすく}}
        \label{fig:controller}
\end{figure*}

In this section, we present a comprehensive controller that encompasses the control of the UAV's body and the tilt angle of the quadlink. The overview of the designed controller is shown in Fig.~\ref{fig:controller}, which is composed of four main components: position controller, attitude controller, control allocation, and tilt angle controller. Among them, we use two types of position controllers; one is for the hovering phase, and the other is for other phases.

In all phases of flight, the controller receives time-series data of the reference forward velocity ${}^{s}v_x$, seen in the stability frame, and the angle of attack $\alpha$. The UAV is then controlled by adjusting rotor thrusts and dynamically changing the tilt angle of the quadlink, $\chi$. 
Note that, as mentioned in Section~\ref{ssec:realize}, the angle of attack $\alpha$ and the pitch angle $\theta$ take the same value when the vehicle moves to the $x$-direction of the inertial frame, as in the case of this paper. Hence, the planned $\alpha$ is utilized as the reference of the pitch angle $\theta$, and we often use $\alpha$ and $\theta$ interchangeably hereafter. 
Also, in this section, we denote the reference value obtained from pre-planned values or a controller as $\ast^\text{ref}$.

\begin{comment}
This section introduces a comprehensive controller that encompasses both rigid body vehicle control and the tilt angle of the quadlink control. The whole controller is shown in Fig.~\ref{fig:controller}. The controller is constructed by combining four distinct components: a Position Controller, an Attitude Controller, a Control  Allocation, and a Tilt Angle Controller.
As mentioned above, the control strategy is shown in Fig.~\ref{fig:transition}.
% VTOL機の制御はホバリング状態とクルーズ状態で違う制御器を定義し，遷移状態に応じて制御器を切り替える．また，遷移中はゲインスケジューリングにより制御するのが広く用いられている．
% 本研究でも同様の制御構造で制御する．
% 本機体の制御はHover, Ready, Transition-Cruiseの4つのフェーズで制御する．
% Hover Phaseは上空に浮上しその場にとどまる．このとき，$\chi=0$であり，すべてのロータは上方向を向いているため，一般的なマルチコプターと同じである．
% Ready Phaseではプランニングした$\alpha,~\chi$の初期値に遷移してその場でホバリングするフェーズである．これにより，次のTransitionにスムーズにつなげる．
% Transitionフェーズは目標の速度になるまで前方へ加速する．加速中は空力の非線形性と機体構造の変化によりモデルは刻一刻と変化するため，同一の制御器で制御することは困難である．そこで状態に応じてあらかじめスケジューリングされたゲインを用いる．最終的に一定速度でのクルーズ状態となり，その時点でのゲインを使ってCruise飛行をする．

Throughout all phases of transition, the controllers receive time-series data of the forward velocity in the stability frame ${}^{s}v_x$ and the angle of attack $\alpha$. The vehicle is then controlled by adjusting rotor thrusts and dynamically changing the parameter $\chi$ in accordance with its current state.
% また，クルーズ飛行では，前章で議論した通り，$\alpha$と$v$を時系列の目標値として与え，状態に応じて$\chi$を変化させることで制御する．
% このとき，$\alpha$に対して厳密な線形化をしてそれぞれを独立して制御することで，時系列の目標値に容易に追従できるようにしている．

\red{この章では右の上付き添え字で$\ast^\text{ref}$とつけたときには実際の値ではなくコントローラで求めた目標値や制御入力値を示すものとする．}    
\end{comment}

\subsection{Position Controller for Hovering Phase}
%First, we design the Position Controller using LQR-optimal control. 
The position controller for the hovering phase utilizes the LQR-optimal control.
The state vector is represented as $\bm \xi = [{}^w\bm x^\top~{}^b\bm v^\top~\bm \eta^\top]^\top$, and the input vector as $\bm u = [{}^{b} f_{\text{rot}z}^\text{ref}~{}^b\bm\omega^{\text{ref}\top}]^\top$.
The reference vector $\bm\xi^\text{ref}$ is denoted as
\begin{align}
    \bm \xi^\text{ref} = \begin{bmatrix}
        0 & 0 & z^\text{ref} & 0 & 0 & 0 & 0 & 0 & 0
    \end{bmatrix}^\top. \label{eq:xiref_hov}
\end{align}
where $z^\text{ref}\in\mbr$ is a reference altitude. %a constant altitude value.
%We obtain the nonlinear state-space equation by transforming Eq.~\eqref{eq:equation_of_motioin} to
By employing $\bm \xi$ and transforming Eq.~\eqref{eq:equation_of_motioin}, the nonlinear state space equation can be obtained as 
\begin{align}
    \dot{\bm\xi} = h(\bm\xi,\bm u).\label{eq:nonlss}
\end{align} 
Note that Eq.~\eqref{eq:nonlss} does not include the aerodynamic forces, as it can be neglected in the hovering phase. %The nonlinear system ABC can be linea
%Note that Eq.~\eqref{eq:nonlss} neglects aerodynamic forces.
The nonlinear system~\eqref{eq:nonlss} can be linearized using a first-order Taylor approximation around the reference equilibrium state Eq.~\eqref{eq:xiref_hov} as
\begin{align}
    \dot{\bm \xi} = \frac{\partial h}{\partial \bm\xi}(\bm\xi\!=\!\bm \xi^\text{ref}, \bm u\!=\!\bm 0) \bm\xi 
                    + \frac{\partial h}{\partial \bm u}(\bm\xi\!=\!\bm \xi^\text{ref}, \bm u\!=\!\bm 0) \bm u.\label{eq:lnss}
\end{align}
%With We then apply LQR-optimal control to determine the feedback gain.
We then apply LQR-optimal control, with the weight matrices $\bm Q_\text{hover}$ and $\bm R_\text{hover}$ for states and inputs, to determine the feedback gain.

\subsection{Position Controller for the Other Phases}

The position controller for other flight phases follows a similar scheme to the one in the hovering phase. Still, some modifications to the position %and the $\chi$
controller are necessitated in order to mitigate unfavorable interference between the pitch angle~$\theta$, which can be regarded as the angle of attack~$\alpha$ as mentioned at the beginning of this section, and the other state variables. 
More specifically, as shown in \eqref{eq:navigation} and \eqref{eq:force}, the pitch angle $\theta$ influences the velocity of the UAV. In the control of a traditional multi-rotor UAV with un-tiltable rotors and no wings, this relationship can be utilized in a way that the UAV accelerates by tilting its body forward. However, in our newly designed UAV, this behavior should be avoided as tilting the body itself forward means heading the nose of the UAV downward, where the UAV cannot generate the lifting force by the aerodynamic force and could fall. 
For this goal, we propose a control method utilizing exact linearization of the pitch angle $\theta$ through coordinate transformation and nonlinear feedback. This approach allows us to control the pitch angle $\theta$ independently from other variables, ensuring $\theta$ follows a pre-planned value. As a result, the proposed scheme favors utilizing $\chi$ rather than $\theta$ to accelerate the UAV's body.
Note that remaining angles such as roll and yaw are necessary for controlling the position and velocity in the $y$-direction. This is because the vehicle does not generate the force in the $y$-direction. Therefore, the system is exactly linearized for the pitch angle only.

To achieve exact linearization of the pitch angle, we introduce a new angle representation, denoted as $\tilde{\bm\eta}=[\tilde{\phi}~\tilde{\theta}~\tilde{\psi}]^\top$, where the vehicle's attitude is expressed in a rotating frame around the Y-X-Z axes with angles $\tilde{\theta}$-$\tilde{\phi}$-$\tilde{\psi}$, in this specific order.
This means the rotational matrix from the inertial frame $\Sigma_w$ to the body frame $\Sigma_b$ becomes $\bm R(\tilde{\bm \eta}) = \bm R_z(\tilde{\psi})\bm R_x(\tilde{\phi})\bm R_y(\tilde{\theta})$.
Let the new velocity vector be $\tilde{\bm v}=\bm R_y(\tilde{\theta}){}^{b}\bm v$, the new force vector be $\tilde{\bm f}_\text{rot}=\bm R_y(\tilde{\theta}){}^{b}\bm f_\text{rot}$, and the new state and the input vector be 
\begin{align}
    \tilde{\bm \xi} =& 
    \begin{bmatrix}
        {}^w\bm x^\top & \tilde{\bm v}^\top & \tilde{\bm \eta}^\top
    \end{bmatrix}^\top,\label{eq;xitilde}\\
    \tilde{\bm u} =&
    \begin{bmatrix}
        \tilde{f}_{\text{rot}x}^\text{ref} & \tilde{f}_{\text{rot}z}^\text{ref} & \dot{\tilde{\bm \eta}}^\top
    \end{bmatrix}^\top.\label{eq:utilde}
\end{align}
% まず，厳密な線形化をするために，慣性座標に対してY-X-Zの順にそれぞれ$\tilde{\theta},~\tilde{\phi},~\tilde{\psi}$回転した新たな姿勢の表現を考える．
% $\bm \tilde{\eta}=[\tilde{\phi}~\tilde{\theta}~\tilde{\psi}]^\top$とし，
% 座標変換した速度を$\tilde{\bm v}^\top = \bm R_y(\tilde{\theta})\bm v^\top$とする．
% 状態ベクトルと入力ベクトルを
% \begin{align}
%     \tilde{\bm \xi} = 
%     \begin{bmatrix}
%         \bm x_e^\top & \tilde{\bm v}^\top & \tilde{\bm \eta}^\top
%     \end{bmatrix}^\top,\label{eq;xitilde}\\
%     \tilde{\bm u} = \begin{bmatrix}
%         {}^{b}\dot{\tilde{v}}_x & {}^{b}\dot{\tilde{v}}_z & \dot{\tilde{\bm \eta}}^\top
%     \end{bmatrix}^\top.\label{eq:utilde}
% \end{align}
From the previously planned value ${}^{s}v_x^\text{ref}$ and $\alpha^\text{ref}$ in Section~\ref{ssec:plan}, the reference vector $\tilde{\bm\xi}^\text{ref}$ is determined as
\begin{align}
    \tilde{\bm \xi}^\text{ref} \!=\! \begin{bmatrix}
        \ast \!&\!\! 0 \!&\!\! z^\text{ref} \!&\!\! {}^{s}v_x^\text{ref}C_{\alpha^\text{ref}} \!&\!\! 0 \!&\!\! {}^{s}v_x^\text{ref}S_{\alpha^\text{ref}} \!&\!\! 0 \!&\!\! \alpha^\text{ref} \!&\!\! 0
    \end{bmatrix}^\top. \label{eq:xitilderef}
\end{align}
%\red{where the $x$ position $\ast$ means to neglect this value.}
Note that $\ast$ at the $x$-element signifies we do not control $x$. Instead, we control the cruise speed~${}^s v_x$.
%The system with only the pitch angle $\tilde{\theta}$ is exactly linearized as follows 
Then, the system is exactly linearized in terms of the pitch angle $\tilde{\theta}$ as
\begin{align}
    \dot{\tilde{\bm\xi}} \!=& \tilde{h}(\tilde{\bm\xi},~\tilde{\bm u})\notag\\
    =&\!\!
    \begin{bmatrix}
        \bm R_z(\tilde{\psi})\bm R_x(\tilde{\phi}) \tilde{\bm v}\\
        {}^{s}\bm f_\text{aero} \!\!+\!\! \bm R_x(\tilde{\phi})^\top\bm R_z(\tilde{\psi})^\top m\bm g \\
        \bm O_3
    \end{bmatrix} \!\!+\!\! 
   \begin{bmatrix}
        \bm O_{3\times 2} \!\!&\! \bm O_3 \\
        \begin{array}{cc}
            1 & 0 \\
            0 & 0 \\
            0 & 1 
        \end{array} \!&\! \bm O_3 \\
        \bm O_{3\times 2} \!\!&\! \bm I_3
    \end{bmatrix} \tilde{\bm u}.\label{eq:exactlinss}
    % \tilde{A} &= 
    % \begin{bmatrix}
    %     ~ & 1 & 0 & 0 & 0 & 0 & - \tilde{v}_x C_{\tilde{\phi}} + \tilde{v}_z S_{\tilde{\phi}} \\
    %     \bm O_3 & 0 & C_{\tilde{\phi}} & -S_{\tilde{\phi}} & -\tilde{v}_x S_{\tilde{\phi}} - \tilde{v}_z C_{\tilde{\phi}} & 0 & \tilde{v}_y \\
    %     ~ & 0 & S_{\tilde{\phi}} & C_{\tilde{\phi}} & -\tilde{v}_x C_{\tilde{\phi}} - \tilde{v}_z S_{\tilde{\phi}} & 0 & 0 \\
    %     ~ & ~ & ~ & ~ & 0 & 0 & -\frac{L}{m} S_{\tilde{\phi}} \\
    %     \bm O_3 & ~ & \bm O_3 & ~ & -(\frac{L}{m}-g)C_{\tilde{\phi}} & 0 & \frac{D}{m} \\
    %     ~ & ~ & ~ & ~ & (\frac{L}{m}-g)S_{\tilde{\phi}} & 0 & 0 \\
    %     \bm O_3 & ~ & \bm O_3 & ~ & ~ & \bm O_3 & ~ 
    % \end{bmatrix},\label{eq:tildeA}\\
    % \tilde{B} &= 
    % \begin{bmatrix}
    %     1 & 0 & ~ \\
    %     0 & 0 & \bm O_3 \\
    %     0 & 1 & ~ \\
    %     ~ & \bm O_{3\times 2} & \bm I_3
    % \end{bmatrix}.\label{eq:tildeB}
\end{align}

Notice that the first six rows in \eqref{eq:exactlinss} do not include $\tilde \theta$, and the right bottom block of the matrix multiplied to $\tilde {\bm u}$ is the identity matrix; hence $\tilde \theta$ only depends on the input $\tilde {\bm u}$ and can independently control it. 
% \red{どこかで空力のピッチ角微分は考えない，外乱としてみなすみたいなことを言う．}
Finally, the control input $\tilde {\bm u}$ for the above model can be transformed in the form of $[f_{\text{rot}x}^\text{ref} ~ f_{\text{rot}z}^\text{ref} ~ {}^b\bm\omega^{\text{ref}\top}]^\top$, which is the same form as in the hovering controller. 

\begin{comment}
from the LQR-controller based on the above model 

Eq.~\eqref{eq:exactlinss}は上6行の並進のダイナミクスに新しいピッチ角$\tilde{\theta}$が入っていないことがわかる．
これは$\tilde{\theta}$が入力のみに線形に依存していることを示しており，このモデルを利用して制御することで$\tilde{\theta}$を並進運動とは独立して制御できる．
\red{どこかで空力のピッチ角微分は考えない，外乱としてみなすみたいなことを言う．}
最後に，上記のモデルを利用して求めた入力を$\tilde{\bm u}\to [f_{\text{rot}x}^\text{ref} ~ f_{\text{rot}x}^\text{ref} ~ \bm\omega^{\text{ref}\top}]^\top$と変換することで，Hoveringコントローラと同じ形式でほかのコントローラに値を渡すことができる．    
\end{comment}

%\red{
Similar to the hovering phase, the nonlinear system \eqref{eq:exactlinss} can be linearized around the equilibrium state \eqref{eq:xitilderef} as
\begin{align}
    \dot{\tilde{\bm \xi}} = \frac{\partial \tilde{h}}{\partial \tilde{\bm \xi}}(\tilde{\bm \xi}\!=\!\tilde{\bm \xi}^\text{ref}, \tilde{\bm u}\!=\!0) \tilde{\bm \xi} 
            + \frac{\partial \tilde{h}}{\partial \tilde{\bm u}}(\tilde{\bm \xi}\!=\!\tilde{\bm \xi}^\text{ref}, \tilde{\bm u}\!=\!0) \tilde{\bm u}.~\label{eq:lnsstilde}
\end{align}
We apply LQR-optimal control, with the weight matrices $\bm Q_\text{cruise}$ and $\bm R_\text{cruise}$ for states and inputs, to determine the feedback gain. The feedback gain of each timestep is calculated previously offline.
%}

% 以上で，$\tilde{\theta}$を厳密に線形化することができた．
% $\tilde{\theta}$と$\alpha$は厳密には違う量であるが，平衡点では一致するためゲインスケジューリングには$\tilde{\theta} = \alpha$を代入する．
% 最終的にHoveringコントローラと同様に局所最適化してLQR制御を適用する．局所最適化の平衡点はSection III.Bで計画された$\bm v, \alpha$に応じてスケジューリングされる．
% % Attitude ControllerはHovering Controllerと同じものを扱い，ゲインのみ調整している．
% % Position controllerとAttitude Controllerの最終的な入力は$[{}^{b}f_x^\text{ref}~0~{}^{b}f_z^\text{ref}~{}^{b}\bm\tau^{\text{ref}\top}]^\top$となる．

% Note that, the controller with $\bm v=0$ is used as the Ready Phase Controller, while with the final value of ${}^{s}v_x$ is used as the Cruise Phase Controller.

\subsection{Attitude Controller}

The attitude controller renders a reference angular acceleration ${}^b\dot{\bm \omega}^\text{ref}$ for a fuselage to make ${}^b\bm\omega$ follow a reference ${}^b\bm \omega^\text{ref}$. We employ a PID controller with P-, I-, and D-gains, $K_P^A$, $K_I^A$, and $K_D^A$, respectively.

%We design the Attitude Controller in such a way that the actual angular velocity $\bm\omega$ follows the desired angular velocity $\bm \omega^\text{ref}$, and we compute the input $\dot{\bm\omega}^\text{ref}$ using a PID controller with feedback $\bm\omega$.

\subsection{Control Allocation}

%With ${}^{b} f_{\text{rot}x}^\text{ref}$, ${}^{b} f_{\text{rot}z}^\text{ref}$, ${}^{b} \bm \tau_\text{rot}^\text{ref}$ calculated by the position and attitude controllers, the control allocation derives a reference $\chi^\text{ref}$ and $\bm f_\text{r+}^\text{ref}$ for achieving those force and torques by finding $\chi$ and $\bm f_\text{r+}$ satisfying the equation \eqref{eq:M_quad}, which is reproduced as 
The control allocation derives a reference $\chi^\text{ref}$ and $\bm f_\text{r+}^\text{ref}$ that achieves ${}^{b} f_{\text{rot}x}^\text{ref}$, ${}^{b} f_{\text{rot}z}^\text{ref}$, ${}^{b} \bm \tau_\text{rot}^\text{ref}=\bm J {}^b\dot{\bm\omega}^\text{ref}$ calculated by the position and attitude controllers. For this goal, the position controller finds $\chi$ and $\bm f_\text{r+}$ satisfying Eq.~\eqref{eq:M_quad}, which is reproduced as 
\begin{align} \label{eq:M_quad_again}
    \begin{bmatrix}
        {}^{b} f_{\text{rot}x}^\text{ref} \\
        {}^{b} f_{\text{rot}z}^\text{ref} \\
        {}^{b} \bm \tau_\text{rot}^\text{ref}
    \end{bmatrix}
    = \bm M(\chi^\text{ref}) \bm f_\text{r+}^\text{ref}.
\end{align}
Note that, in the hovering phase, we substitute $\chi^\text{ref}=0$ and $f_{\text{rot}x}^\text{ref}=0$ into \eqref{eq:M_quad_again} before obtaining $\bm f_\text{r+}^\text{ref}$.

Let us emphasize that the reference thrust $\bm f_\text{r+}^\text{ref}$ only designates the thrust of two rotors at the tail and the summations of two rotors located on the same side of the quadlink. In other words, $\bm f_\text{r+}^\text{ref}$ does not specify the thrust of each rotor mounted on the quadlink, namely $f_i,~i=\{1 \cdots 4 \}$. These values are determined by mixing $\bm f_\text{r+}^\text{ref}$ and $\bm f_\text{r-}^\text{ref}$, as explained in the next subsection.

\begin{comment}
Section2.Cの議論の通りEq.~\eqref{eq:M_quad}は，$\chi$を動かすことで厳密に$[{}^{b}f_{\text{rot}x}~{}^{b}f_{\text{rot}z}~{}^{b}\bm\tau_\text{rot}^\top]^\top\in\mbr^5\to(\chi,~\bm f_\text{r+})\in\mbr^5$の逆変換をすることができる．
その値は下記の式を解くことで得ることができる．
\begin{align}
    (\chi^\text{ref},~\bm f_\text{r+}^\text{ref}),~\mathrm{s.t.}~
    \begin{bmatrix}
        {}^{b} f_{\text{rot}x}^\text{ref} \\
        {}^{b} f_{\text{rot}z}^\text{ref} \\
        {}^{b} \bm \tau_\text{rot}^\text{ref}
    \end{bmatrix}
    = \bm M(\chi^\text{ref}) \bm f_\text{r+}^\text{ref}
\end{align}
Hoveringのときは$\chi^\text{ref}=0,~f_{\text{rot}x}^\text{ref}=0$を代入して次式によりAllocationをしている．
\begin{align}
    &\chi^\text{ref}=0, \\
    &\bm f_\text{r+}^\text{ref},~\mathrm{s.t.}~
    \begin{bmatrix}
        {}^{b} f_{\text{rot}z}^\text{ref} \\
        {}^{b} \bm \tau_\text{rot}^\text{ref}
    \end{bmatrix}
    = \bm M(\chi=0)_{\neg x} \bm f_\text{r+}^\text{ref}
\end{align}
ここで$\bm M_{\neg x}$は$\bm M$から1行目を除いた行列である．
この行列は$\mbr^4\times\mbr^4$であるため，$\bm f_\text{r+}^\text{ref}$は逆行列を使って次式のように求めることができる．
\begin{align}
    \bm f_\text{r+}^\text{ref}
    = \bm M(0)_{\neg x}^{-1} \begin{bmatrix}
        {}^{b} f_{\text{rot}z}^\text{ref} \\
        {}^{b} \bm \tau_\text{rot}^\text{ref}
    \end{bmatrix}
\end{align}
上記で求めた$\chi^\text{ref}$を次のTilt Angle Controllerに渡す．
\end{comment}

\subsection{Tilt Angle Controller}

The tilt angle controller calculates $\bm f_\text{r-}= [f_1\!-\!f_2~f_3\!-\!f_4]^\top$, which makes $\chi$ follows a reference value~$\chi^\text{ref}$. Since $\bm f_\text{r-}^\text{ref}$ is the difference of thrust between the front and rear rotors in the quadlink, combining $\bm f_\text{r+}$ and $\bm f_\text{r-}$ via a mixer determines each rotor's thrust in the quadlink, as mentioned in Remark~\ref{rem:quadlink}. We employ a PID controller to obtain the reference torque input $\tau_\chi^\text{ref}$, which is utilized to yield $\bm f_\text{r-}^\text{ref}$ as 
%制御器の目標推力入力から$\chi^\text{ref}$を求め，独立した$\chi$コントローラでPID制御する．
%PID制御でリンクトルク目標値$\tau_\chi^\text{ref}$を求めたあと，Eq.~\eqref{eq:Mchi_quad}の逆変換をすることで$\bm f_\text{r-}^\text{ref}$を求める．
\begin{align}
    \bm f_\text{r-}^\text{ref} =  
    \begin{bmatrix}
        l_{fh} & l_{fh}
    \end{bmatrix}^\dagger
    \tau_\chi^\text{ref},
\end{align}
where $\dagger$ denotes the Moore-Penrose inverse.
P-, I-, and D-gains are $K_P^\chi$, $K_I^\chi$, and $K_D^\chi$, respectively.
\begin{comment}
ここで$\dagger$はMoore–Penrose inverseである．
最終的に，Control Allocationで得た$\bm f_\text{r+}^\text{ref}$と合わせて$\bm f_\text{rot}^\text{ref}$を得る．
\begin{align}
    \begin{bmatrix}
        \bm f_\text{r+}^\text{ref} \\
        \bm f_\text{r-}^\text{ref}
    \end{bmatrix} \to \bm f_\text{rot}^\text{ref}
\end{align}

\red{mixerについて，remarkとも関係するのでまだかけていない}    
\end{comment}

% Eq.~\eqref{eq:M_quad}はランクが4であり，$[{}^{b}f^\text{ref}_{\text{rot}x}~0~{}^{b}f^\text{ref}_{\text{rot}z}~{}^{b}\bm\tau_\text{rot}]^\top \to \bm f_\text{r+}$の逆変換が一般には存在しない．
% しかし，$\bm P[{}^{b}f^\text{ref}_{\text{rot}x}~0~{}^{b}f^\text{ref}_{\text{rot}z}~{}^{b}\bm\tau_\text{rot}]=[0~0~{}^{b}f^\text{ref}_{\text{rot}\ast}~{}^{b}\bm\tau_\text{rot}]$となる変換に対して，
% $\bm P\bm M_\text{+}(\chi)=[\bm O_{2\times 6}~\tilde{\bm M}_\text{+}^\top]^\top$となる$\chi$が見つけられれば．
% \begin{align}
%     \begin{bmatrix}
%         {}^{b}f^\text{ref}_{\text{rot}\ast} \\
%         {}^{b}\bm\tau_\text{rot}
%     \end{bmatrix}
%     = \tilde{\bm M}_\text{+} \bm f_\text{rot}
% \end{align}
% という$\mbr^4\leftrightarrow\mbr^4$の全単射の変換となり，Allocationができるようになる．
% この$\chi$は厳密に求めることができ，次の$\chi$ Controllerの目標値として使用する．

% \documentclass[ieeeconf]{subfiles}
% \begin{document}

\section{Simulation}

This section presents simulation study, where we demonstrate the proposed control strategy successfully achieves a transition from hovering to cruise flight mode. The simulation also verifies that the proposed method converges to the desired cruising speed, the angle of attack, and the tilt angle of the quadlink specified by Theorem 1 in the cruise flight mode. 
The parameters of the vehicle are set as
$m\!=\!0.5~\mathrm{[kg]}$, 
$\bm J\!=\!\mathrm{diag}(0.03, 0.05, 0.05)~\mathrm{[kgm^2]}$, and 
$l_{f}\!=\!l_{fw}\!=\!l_{fh}\!=\!l_{b}\!=\!l_{bw}\!=\!0.1~\mathrm{[m]}$, with setting 
the inertial moment of the quadlink as $0.01~\mathrm{[kgm^2]}$.
Also, the control gains are 
$\bm Q_\text{hover}=\mathrm{diag}(1,1,10,1,1,1,1,1,1)$, 
$\bm R_\text{hover}=\mathrm{diag}(1,2,2,1)\times 10^2$, 
$\bm Q_\text{cruise}=\mathrm{diag}(1,1,10^2,10^4,1,10^4,1,10^3,1)$, 
$\bm R_\text{cruise}=\mathrm{diag}(31,31,2,2,2)\times 10^2$, 
$[K_P^A,~K_I^A,~K_D^A]=[100,1000,0.3]$, and 
$[K_P^\chi,~K_I^\chi,~K_D^\chi]=[5,1,4]$. %Also, the inertial moment of the quadlink is set as $0.01~\mathrm{[kgm^2]}$.

By denoting the simulation time as $t$, the performed simulation can be divided into the following three phases: 
\begin{itemize}
    \item From the initial $z$-position $z=0$, the UAV moves to the desired $z$-position $z = -5$ (altitude of $5$\,m) and keep hovering until $t = 20$\,s,
    \item the UAV is in the ready phase from $t = 20$\,s to $t = 30$\,s, where $\alpha$ and $\chi$ converge the desired value at $t = 25$\,s and hover until $t = 30$\,s with fixing $\alpha$ and $\chi$, and
    \item the UAV is in the acceleration phase from $t = 30$\,s to $t=50$\,s, and enters the cruise phase with ${}^w v_x = 20$\,m/s until $t=80$\,s.
\end{itemize}
Note that, in all phases, the UAV's reference $y$-position is $y=0$. 
The above reference trajectories of ${}^w v_x$, ${}^wy$, and ${}^wz$ are depicted in black dashed lines in Figs.~\ref{fig:result}(\subref{fig:result_vx}) and~\ref{fig:result}(\subref{fig:result_yz}).

% \red{以前，$\alpha = \theta$と言っていた気がするのですが，加速Phaseでは$z$がズレてしまっている（z方向の速度がでているので，body frameとstability frameが一致しない）ので，$\alpha \neq \theta$になってしまっているのではという懸念があります}
% \violet{これFig.~\ref{fig:result}(\subref{fig:result_vyz})を見ると速度は0.1m/s程度なのでx方向の速度がある程度出たら無視できると主張してもいいかもしれないですね．加速初めは無視できなさそうですが…}

The evolution of all states of the UAV during simulation is shown in Fig.~\ref{fig:result}. 
All the values, except for the body angular velocity in Fig. 8(f), are seen in the inertial frame.
% The body angular velocity in the body frame is shown in Fig.~\ref{fig:result}(\subref{fig:result_omega}), and the other values are seen in the inertial frame.
%シミュレーションの結果をFig.~\ref{fig:result}に示す．
%結果はすべて慣性座標から見たものである．
Fig.~\ref{fig:result}(\subref{fig:result_vx}) depicts the cruising speed, and Fig.~\ref{fig:result}(\subref{fig:result_eta}) shows the roll, pitch, and yaw angle. 
% \red{Note that since the velocities in $y$ and $z$ can be negligible in the acceleration and cruise phase, we can regard $\theta = \alpha$, as mentioned in Section~\ref{ssec:realize}. } 
% \violet{section2で言ってる？}
Also, the evolution of $\chi$ is illustrated in Fig.~\ref{fig:result}(\subref{fig:result_chi}).

Let us investigate the evolution of ${^s v_x}$, $\theta$, and $\chi$, shown in Figs.~\ref{fig:result}(\subref{fig:result_vx}), \ref{fig:result}(\subref{fig:result_eta}), and \ref{fig:result}(\subref{fig:result_chi}). The black dashed line in Fig.~\ref{fig:result}(\subref{fig:result_eta}) and the orange dashed line in Fig.~\ref{fig:result}(\subref{fig:result_chi}) correspond with the planned value in Fig.~\ref{fig:equilibrium}, calculated from Theorem~\ref{thm:cruise}. Let us emphasize that, as mentioned in Sections~\ref{sec:cruisability} and \ref{sec:controller}, this planned value corresponds with the equilibrium points, and hence, during the acceleration phase, the tilt angle of the quadlink has to deviate from the path in order to accelerate or decelerate, as detailed soon. 
First, in the hovering phase, although a deviation of $\theta$ can be confirmed before the UAV converges to $z = -5$\,m, its effect on the cruising speed ${^s v_x}$ and $y, z$ positions is small and can be negligible. 
Second, in the ready phase, ${^s v_x}$, $\theta$, and $\chi$ follow the planned values, and the UAV achieves static hovering, i.e., its position does not change, as shown in Fig.~\ref{fig:result}(\subref{fig:result_yz}). % \ref{fig:result}(\subref{fig:result_vx}) and 
In the acceleration phase, as mentioned earlier, the proposed control design tilts the quadlink forward than the path generated from the equilibrium conditions to accelerate, as in Fig.~\ref{fig:result}(\subref{fig:result_chi}). 
Notably, the cruise speed ${^s v_x}$ and the pitch angle $\theta$, corresponding to the angle of attack $\alpha$, do not deviate from the planned path, hence restraining the change of the aerodynamic force from the planned one. This strategy alleviates the complicated effect of the aerodynamic force. 
During this acceleration phase, $z$-position shows slight changes from the reference. This deviation can be regarded as a price of increasing the cruise speed because tilting quadlink forward leads to a decrease of the force canceling the gravity. 
%This deviation can be regarded as a price of increasing the cruise speed because tilting quadlink forward leads to a decrease of the force canceling the gravity. 
Nevertheless, as the cruising speed increases, this deviation is eliminated. 
Lastly, in the cruise phase, all the state of the UAV converges the planned one, namely the equilibrium point, and achieving the cruise flight with the desired cruise speed. 
The above results demonstrate the proposed control strategy successfully achieve the transitions of flight mode, while achieving the cruise flight with the desired cruising speed.

\begin{figure}[t]
        % \centering
        % \includegraphics[width=\linewidth]{"figure/simulation/snapshot.eps"}
        % % \includegraphics[width=\linewidth]{"figure/simulation/snapshot_noangle.eps"}
        % \caption{snapshot\red{真横から見たのと角度つけてプロットしたものとコメントで二種類切り替えられます}}
        % \label{fig:snapshot}
    \begin{minipage}{0.49\linewidth}
        \centering
        \includegraphics[trim = 0cm 2.5cm 0cm 2.5cm, clip=true,width=\linewidth]{"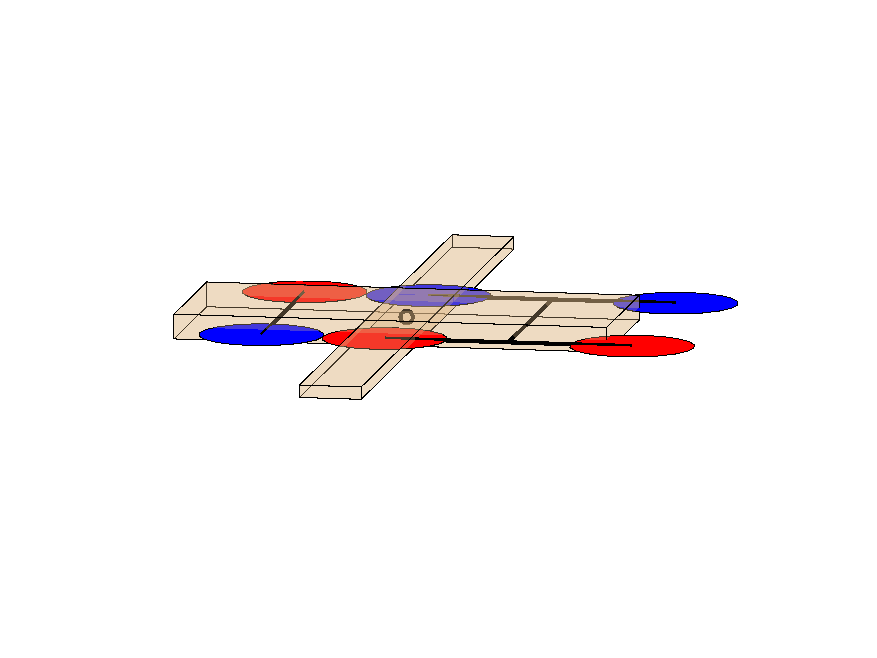"}
        \subcaption{$t=0$\,s}
        \label{fig:snapshot0}
    \end{minipage}
    \begin{minipage}{0.49\linewidth}
        \centering
        \includegraphics[trim = 0cm 2.5cm 0cm 2.5cm, clip=true,width=\linewidth]{"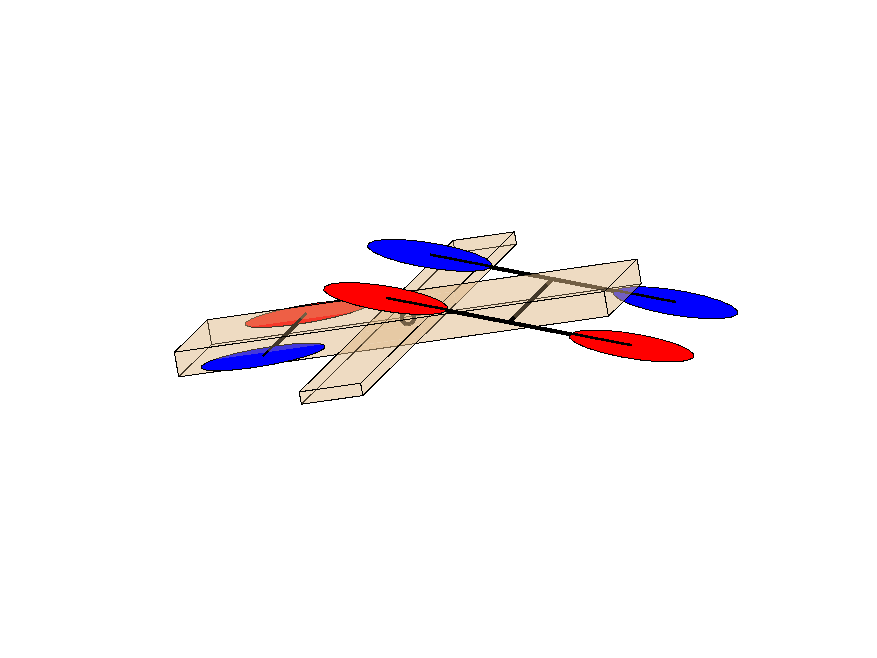"}
        \subcaption{$t=30$\,s}
        \label{fig:snapshot30}
    \end{minipage}
    \begin{minipage}{0.49\linewidth}
        \centering
        \includegraphics[trim = 0cm 1.5cm 0cm 1cm, clip=true,width=\linewidth]{"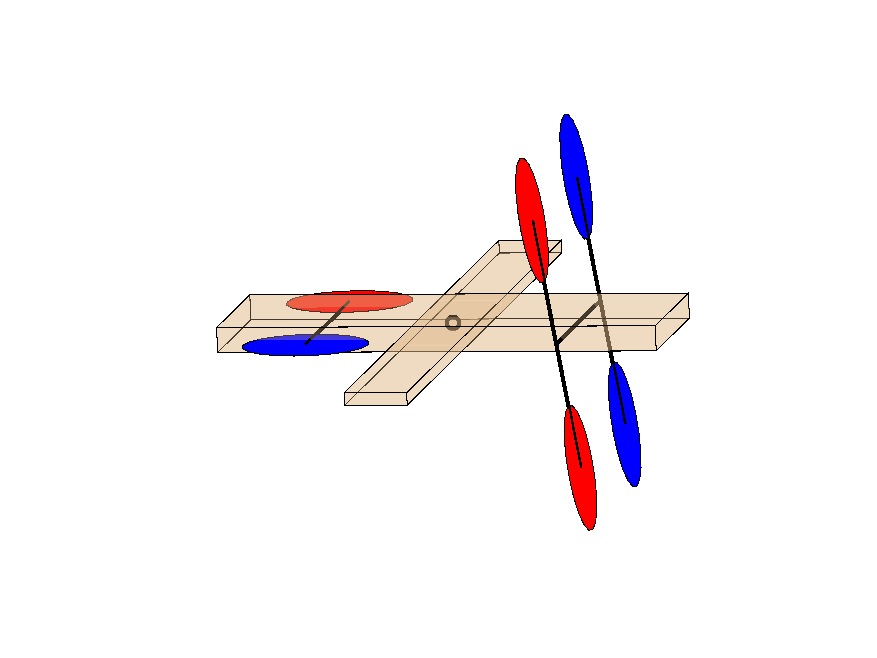"}
        \subcaption{$t=50$\,s}
        \label{fig:snapshot50}
    \end{minipage}
    \begin{minipage}{0.49\linewidth}
        \centering
        \includegraphics[trim = 0cm 1.5cm 0cm 1cm, clip=true,width=\linewidth]{"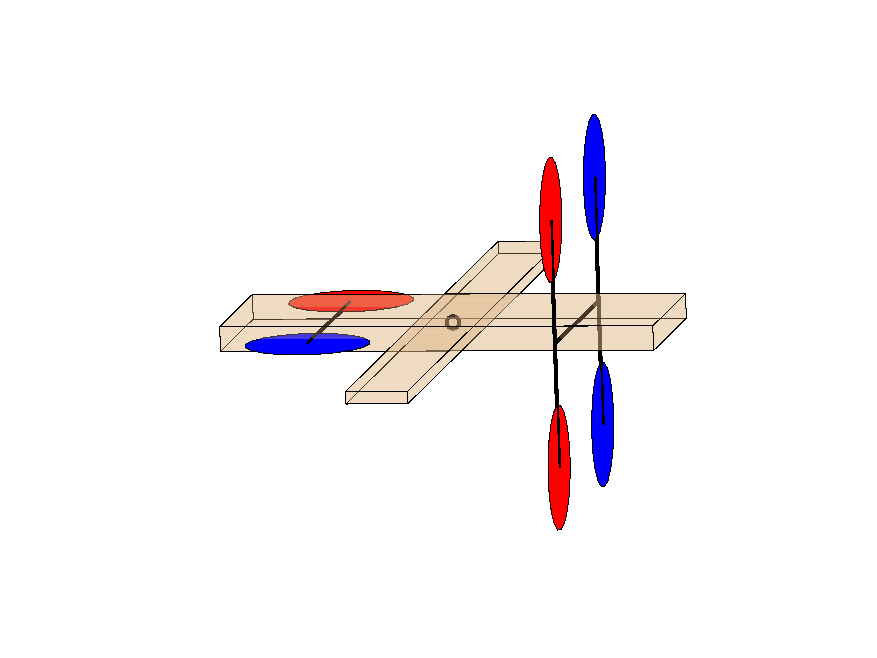"}
        \subcaption{$t=80$\,s}
        \label{fig:snapshot80}
    \end{minipage}
    \caption{Snapshots illustrating attitude and $\chi$ of the UAV, where the UAV moves from left to right.}
    \label{fig:snapshot}
\end{figure}

\begin{figure}[t]
    % \begin{minipage}{0.48\linewidth}
    %     \centering
    %     \includegraphics[width=\linewidth]{"figure/simulation/chi_val_control_posx.eps"}
    %     \subcaption{x}
    %     \label{fig:result_x} 
    % \end{minipage}
        \begin{minipage}{0.49\linewidth}
        \centering
        \includegraphics[width=\linewidth]{"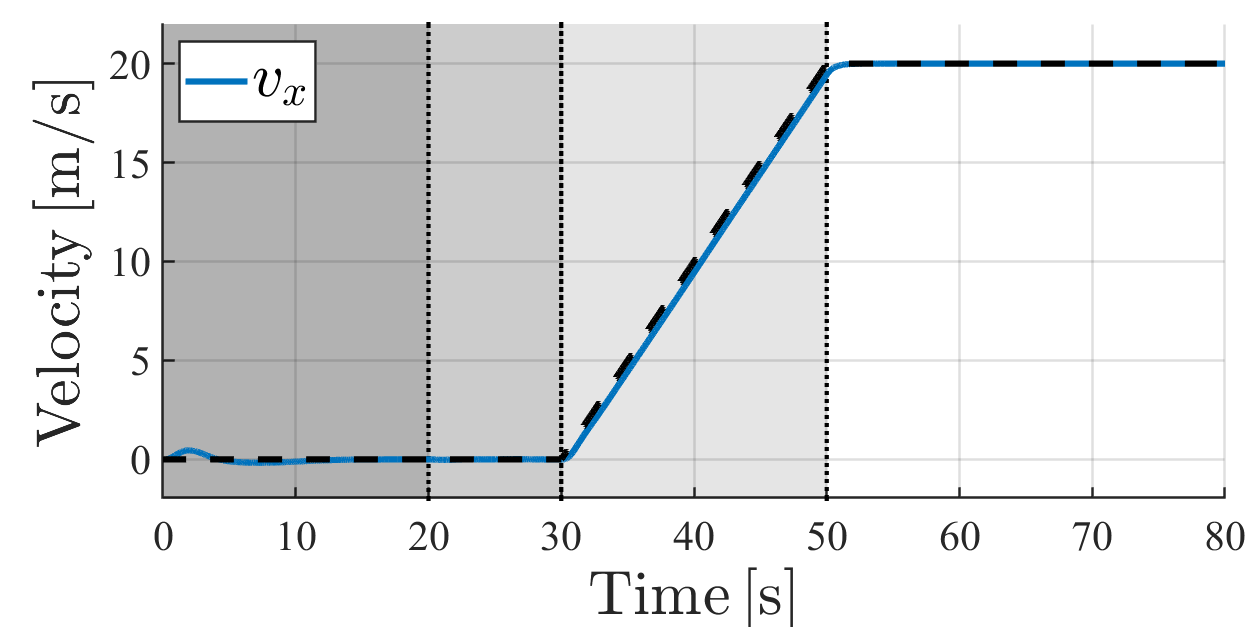"}
        \subcaption{${}^w v_x$}
        \label{fig:result_vx} 
    \end{minipage}
    \begin{minipage}{0.49\linewidth}
        \centering
        \includegraphics[width=\linewidth]{"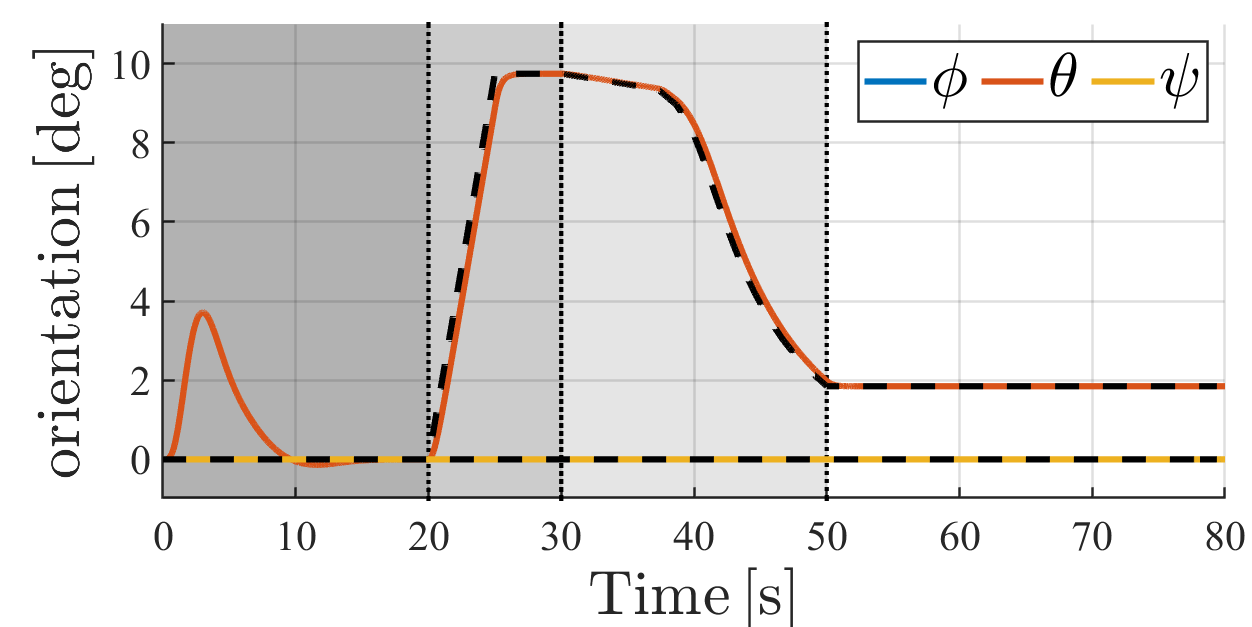"}
        \subcaption{$\bm\eta$}
        \label{fig:result_eta}
    \end{minipage}
    \begin{minipage}{0.49\linewidth}
        \centering
        \includegraphics[width=\linewidth]{"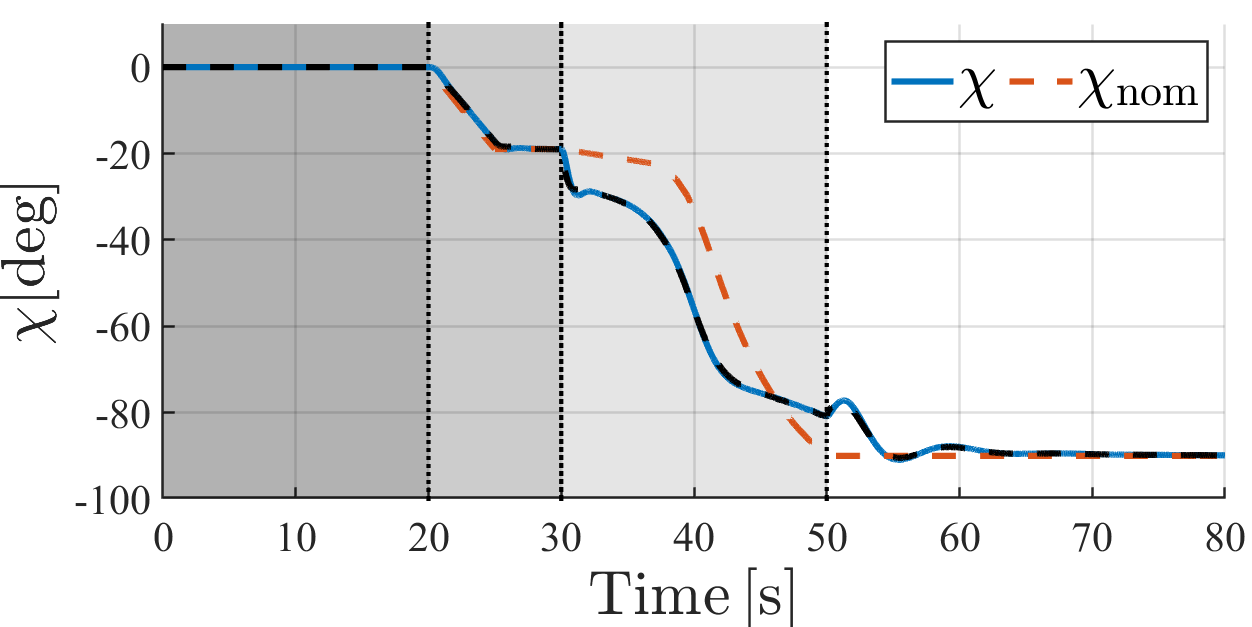"}
        \subcaption{$\chi$}
        \label{fig:result_chi}
    \end{minipage}
    \begin{minipage}{0.49\linewidth}
        \centering
        \includegraphics[width=\linewidth]{"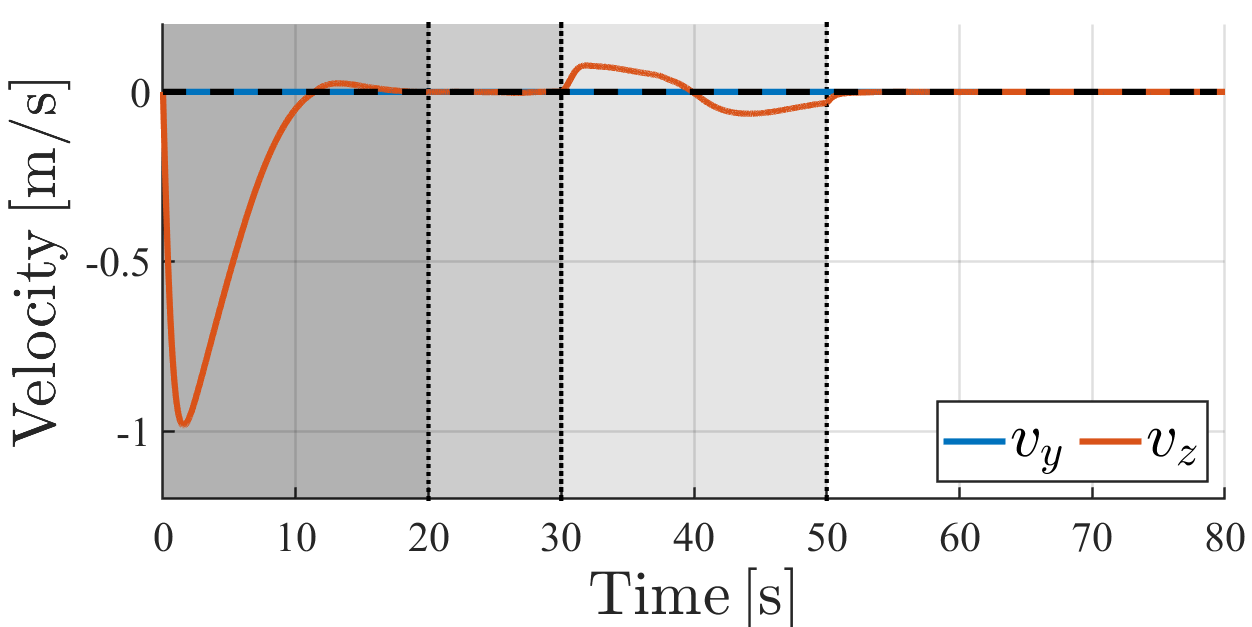"}
        \subcaption{${}^w v_y,~{}^w v_z$}
        \label{fig:result_vyz}
    \end{minipage}
    \begin{minipage}{0.49\linewidth}
        \centering
        \includegraphics[width=\linewidth]{"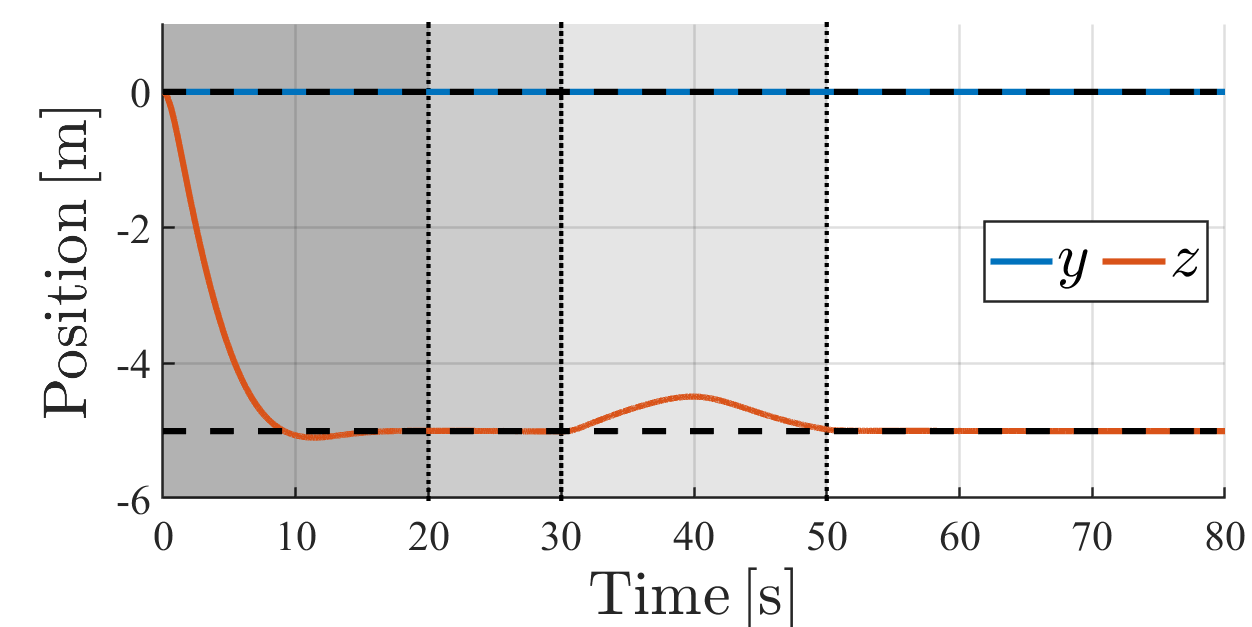"}
        \subcaption{${}^w y,{}^w z$}
        \label{fig:result_yz}
    \end{minipage}
    \begin{minipage}{0.49\linewidth}
        \centering
        \includegraphics[width=\linewidth]{"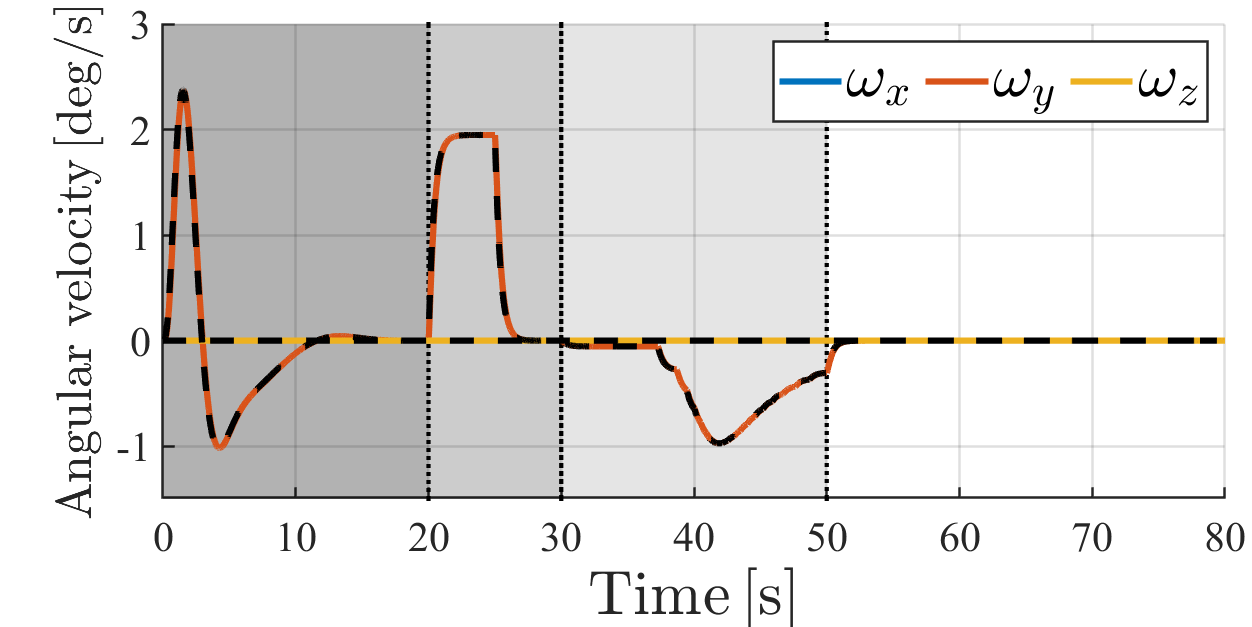"}
        \subcaption{${}^{b}\bm\omega$}
        \label{fig:result_omega}
    \end{minipage}
    % \begin{minipage}{0.48\linewidth}
    %     \centering
    %     \includegraphics[width=\linewidth]{"figure/simulation/chi_val_control_rotor.eps"}
    %     \subcaption{rotor thrust}
    %     \label{fig:result_rotor}
    % \end{minipage}
    \caption{Simulation results.}
    \label{fig:result}
\end{figure}

% \end{document}
% \documentclass[ieeeconf]{subfiles}
% \begin{document}

\section{CONCLUSION}
This paper presented a novel VTOL UAV having six rotors, four of which are mounted to a tiltable link attached to the fuselage via a passive joint. We first presented the flight modes for the designed UAV while describing the overall strategy of manipulating the angle of attack and the tilt angle of the quadlink. We derived the dynamical model of the UAV and analyzed its controllability. The derived condition for the UAV's controllability is utilized for planning the angle of attack and the tilt angle of the quadlink. 
We then designed the controller that leverages the tiltable quadlink to accelerate the UAV while the deviations in the angle of attack from the equilibrium value are suppressed to avoid the changes of the aerodynamic force. The validity of the designed UAV and the proposed controller are demonstrated in simulation studies.
% \end{document}

\bibliography{IEEEexample}

\begin{thebibliography}{10}
\providecommand{\url}[1]{#1}
\csname url@rmstyle\endcsname
\providecommand{\newblock}{\relax}
\providecommand{\bibinfo}[2]{#2}
\providecommand\BIBentrySTDinterwordspacing{\spaceskip=0pt\relax}
\providecommand\BIBentryALTinterwordstretchfactor{4}
\providecommand\BIBentryALTinterwordspacing{\spaceskip=\fontdimen2\font plus
\BIBentryALTinterwordstretchfactor\fontdimen3\font minus
  \fontdimen4\font\relax}
\providecommand\BIBforeignlanguage[2]{{%
\expandafter\ifx\csname l@#1\endcsname\relax
\typeout{** WARNING: IEEEtran.bst: No hyphenation pattern has been}%
\typeout{** loaded for the language `#1'. Using the pattern for}%
\typeout{** the default language instead.}%
\else
\language=\csname l@#1\endcsname
\fi
#2}}

\bibitem{hatanaka_agriculture}
M.~Mammarella, C.~Donati, T.~Shimizu, M.~Suenaga, L.~Comba, A.~Biglia, K.~Uto,
  T.~Hatanaka, P.~Gay, and F.~Dabbene, ``{3D} map reconstruction of an orchard
  using an angle-aware covering control strategy,'' \emph{IFAC-PapersOnLine},
  vol.~55, no.~32, pp. 271--276, 2022.

\bibitem{Sekiguchi202190}
W.~Eikyu, K.~Sekiguchi, and K.~Nonaka, ``Nonlinear control for the extended
  model of the load-suspended {UAV} based on the experiments,''
  \emph{IFAC-PapersOnLine}, vol.~54, no.~14, pp. 90--95, 2021.

\bibitem{Erdelj2017_SAR}
M.~Erdelj, E.~Natalizio, K.~R. Chowdhury, and I.~F. Akyildiz, ``Help from the
  sky: Leveraging {UAV}s for disaster management,'' \emph{IEEE Pervasive
  Computing}, vol.~16, no.~1, pp. 24--32, 2017.

\bibitem{Boon17}
M.~A. Boon, A.~P. Drijfhout, and S.~Tesfamichael, ``Comparison of a fixed-wing
  and multi-rotor {UAV} for environmental mapping applications: {A} case
  study,'' \emph{Int. Arch. Photogrammetry, Remote Sensing and Spatial
  Information Sci.}, vol. XLII-2/W6, pp. 47--54, 2017.

\bibitem{saeed_survey_2018}
A.~S. Saeed, A.~B. Younes, C.~Cai, and G.~Cai, ``A survey of hybrid unmanned
  aerial vehicles,'' \emph{Prog. Aerospace Sci.}, vol.~98, pp. 91--105, 2018.

\bibitem{Kohno2014}
S.~Kohno and K.~Uchiyama, ``Design of robust controller of fixed-wing {UAV} for
  transition flight,'' in \emph{Int. Conf. on Unmanned Aircraft Syst.}, 2014,
  pp. 1111--1116.

\bibitem{9051852}
B.~Li, J.~Sun, W.~Zhou, C.-Y. Wen, K.~H. Low, and C.-K. Chen, ``Transition
  optimization for a {VTOL} tail-sitter {UAV},'' \emph{IEEE/ASME Trans.
  Mechatronics}, vol.~25, no.~5, pp. 2534--2545, 2020.

\bibitem{Tal2023_tail_sitter}
E.~Tal, G.~Ryou, and S.~Karaman, ``Aerobatic trajectory generation for a {VTOL}
  fixed-wing aircraft using differential flatness,'' \emph{IEEE Trans. Robot.},
  pp. 1--15, 2023.

\bibitem{Kikumoto2022}
C.~Kikumoto, T.~Urakubo, K.~Sabe, and Y.~Hazama, ``Back-transition control with
  large deceleration for a dual propulsion {VTOL} {UAV} based on its
  maneuverability,'' \emph{IEEE Robot. Autom. Lett.}, vol.~7, no.~4, pp.
  11\,697--11\,704, 2022.

\bibitem{LIU2017135}
Z.~Liu, Y.~He, L.~Yang, and J.~Han, ``Control techniques of tilt rotor unmanned
  aerial vehicle systems: A review,'' \emph{Chinese J. Aeronautics}, vol.~30,
  no.~1, pp. 135--148, 2017.

\bibitem{7152383}
R.~G. Hern\'{a}ndez-Garc\'{i}a and H.~Rodr\'{i}guez-Cort\'{e}s, ``Transition
  flight control of a cyclic tiltrotor {UAV} based on the gain-scheduling
  strategy,'' in \emph{Int. Conf. Unmanned Aircraft Syst.}, 2015, pp. 951--956.

\bibitem{hernandez-garcia_total_2013}
------, ``A total energy control system design for the transition phase of a
  tiltrotor aerial vehicle,'' \emph{IFAC Proc. Vol.}, vol.~46, no.~30, pp.
  52--57, 2013.

\bibitem{6300840}
X.~Fang, Q.~Lin, Y.~Wang, and L.~Zheng, ``Control strategy design for the
  transitional mode of tiltrotor {UAV},'' in \emph{IEEE Int. Conf. Industrial
  Informatics}, 2012, pp. 248--253.

\bibitem{9444145}
D.~Rohr, M.~Studiger, T.~Stastny, N.~R.~J. Lawrance, and R.~Siegwart,
  ``Nonlinear model predictive velocity control of a {VTOL} tiltwing {UAV},''
  \emph{IEEE Robot. Autom. Lett.}, vol.~6, no.~3, pp. 5776--5783, 2021.

\bibitem{9183353}
M.~Allenspach and G.~J. Ducard, ``Model predictive control of a convertible
  tiltrotor unmanned aerial vehicle,'' in \emph{Mediterranean Conf. Control and
  Autom.}, 2020, pp. 715--720.

\bibitem{mousaei_design_2022}
M.~Mousaei, J.~Geng, A.~Keipour, D.~Bai, and S.~Scherer, ``Design, modeling and
  control for a tilt-rotor {VTOL} {UAV} in the presence of actuator failure,''
  in \emph{IEEE/RSJ Int. Conf. Intelligent Robots and Syst.}, 2022, pp.
  4310--4317.

\bibitem{chiappinelli_passive}
R.~Chiappinelli, M.~Cohen, M.~Doff-Sotta, M.~Nahon, J.~R. Forbes, and
  J.~Apkarian, ``Modeling and control of a passively-coupled tilt-rotor
  vertical takeoff and landing aircraft,'' in \emph{2019 International
  Conference on Robotics and Automation}, 2019, pp. 4141--4147.

\bibitem{9393789}
S.~Mochida, R.~Matsuda, T.~Ibuki, and M.~Sampei, ``A geometric method of
  hoverability analysis for multirotor {UAV}s with upward-oriented rotors,''
  \emph{IEEE Trans. Robot.}, vol.~37, no.~5, pp. 1765--1779, 2021.

\bibitem{Hatanaka2015_Springer}
T.~Hatanaka, N.~Chopra, M.~Fujita, and M.~W. Spong, \emph{{Passivity-Based
  Control and Estimation in Networked Robot.}}\hskip 1em plus 0.5em minus
  0.4em\relax Springer, 2015.

\bibitem{ozdemir_design_2014}
U.~Ozdemir, Y.~O. Aktas, A.~Vuruskan, Y.~Dereli, A.~F. Tarhan, K.~Demirbag,
  A.~Erdem, G.~D. Kalaycioglu, I.~Ozkol, and G.~Inalhan,
  ``\BIBforeignlanguage{en}{Design of a commercial hybrid {VTOL} {UAV}
  system},'' \emph{\BIBforeignlanguage{en}{J. Intelligent \& Robot. Syst.}},
  vol.~74, no.~1, pp. 371--393, Apr. 2014.

\end{thebibliography}


\begin{thebibliography}{99}

\bibitem{c1} G. O. Young, Synthetic structure of industrial plastics (Book style with paper title and editor), 	in Plastics, 2nd ed. vol. 3, J. Peters, Ed.  New York: McGraw-Hill, 1964, pp. 1564.
\bibitem{c2} W.-K. Chen, Linear Networks and Systems (Book style).	Belmont, CA: Wadsworth, 1993, pp. 123135.
\bibitem{c3} H. Poor, An Introduction to Signal Detection and Estimation.   New York: Springer-Verlag, 1985, ch. 4.
\bibitem{c4} B. Smith, An approach to graphs of linear forms (Unpublished work style), unpublished.
\bibitem{c5} E. H. Miller, A note on reflector arrays (Periodical styleAccepted for publication), IEEE Trans. Antennas Propagat., to be publised.
\bibitem{c6} J. Wang, Fundamentals of erbium-doped fiber amplifiers arrays (Periodical styleSubmitted for publication), IEEE J. Quantum Electron., submitted for publication.
\bibitem{c7} C. J. Kaufman, Rocky Mountain Research Lab., Boulder, CO, private communication, May 1995.
\bibitem{c8} Y. Yorozu, M. Hirano, K. Oka, and Y. Tagawa, Electron spectroscopy studies on magneto-optical media and plastic substrate interfaces(Translation Journals style), IEEE Transl. J. Magn.Jpn., vol. 2, Aug. 1987, pp. 740741 [Dig. 9th Annu. Conf. Magnetics Japan, 1982, p. 301].
\bibitem{c9} M. Young, The Techincal Writers Handbook.  Mill Valley, CA: University Science, 1989.
\bibitem{c10} J. U. Duncombe, Infrared navigationPart I: An assessment of feasibility (Periodical style), IEEE Trans. Electron Devices, vol. ED-11, pp. 3439, Jan. 1959.
\bibitem{c11} S. Chen, B. Mulgrew, and P. M. Grant, A clustering technique for digital communications channel equalization using radial basis function networks, IEEE Trans. Neural Networks, vol. 4, pp. 570578, July 1993.
\bibitem{c12} R. W. Lucky, Automatic equalization for digital communication, Bell Syst. Tech. J., vol. 44, no. 4, pp. 547588, Apr. 1965.
\bibitem{c13} S. P. Bingulac, On the compatibility of adaptive controllers (Published Conference Proceedings style), in Proc. 4th Annu. Allerton Conf. Circuits and Systems Theory, New York, 1994, pp. 816.
\bibitem{c14} G. R. Faulhaber, Design of service systems with priority reservation, in Conf. Rec. 1995 IEEE Int. Conf. Communications, pp. 38.
\bibitem{c15} W. D. Doyle, Magnetization reversal in films with biaxial anisotropy, in 1987 Proc. INTERMAG Conf., pp. 2.2-12.2-6.
\bibitem{c16} G. W. Juette and L. E. Zeffanella, Radio noise currents n short sections on bundle conductors (Presented Conference Paper style), presented at the IEEE Summer power Meeting, Dallas, TX, June 2227, 1990, Paper 90 SM 690-0 PWRS.
\bibitem{c17} J. G. Kreifeldt, An analysis of surface-detected EMG as an amplitude-modulated noise, presented at the 1989 Int. Conf. Medicine and Biological Engineering, Chicago, IL.
\bibitem{c18} J. Williams, Narrow-band analyzer (Thesis or Dissertation style), Ph.D. dissertation, Dept. Elect. Eng., Harvard Univ., Cambridge, MA, 1993. 
\bibitem{c19} N. Kawasaki, Parametric study of thermal and chemical nonequilibrium nozzle flow, M.S. thesis, Dept. Electron. Eng., Osaka Univ., Osaka, Japan, 1993.
\bibitem{c20} J. P. Wilkinson, Nonlinear resonant circuit devices (Patent style), U.S. Patent 3 624 12, July 16, 1990. 






\end{thebibliography}
\bibliographystyle{IEEEtran}

\end{document}